\documentstyle[aps,twocolumn,epsf,psfig]{revtex}
\begin{document}

\def\a{{\bf{a}}}
\def\r{{\bf{r}}}
\def\k{{\bf{k}}}
\def\q{{\bf{q}}}
\def\p{{\bf{p}}}
\def\K{{\bf{K}}}
\def\Q{{\bf{Q}}}
\def\P{{\bf{P}}}
\def\M{{\bf M}}
\def\kt{{\tilde{\k}}}
\def\Gbar{\bar{G}}
\def\Ghat{\hat{G}}
\def\Ghatc{\hat{G}_c}
\def\Gc{{G_c}}
\def\bGc{ {{\bf{G}}_c} }
\def\chibar{\bar{\chi}}
\def\chic{{\chi_c}}
\def\Sigmac{{\Sigma_c}}
\def\Phic{{\Phi_c}}
\def\Gscript{{\cal{G}}}
\def\Gammac{{\Gamma_c}}
\def\Dk{\Delta k}
\def\grad{\nabla}
\def\sign{{\rm{sign}}}

\def\r{{\bf{r}}}
\def\R{{\bf{R}}}
\def\i{{\bf{i}}}
\def\j{{\bf{j}}}
\def\m{{\bf{m}}}
\def\k{{\bf{k}}}
\def\K{{\bf{K}}}
\def\P{{\bf{P}}}
\def\q{{\bf{q}}}
\def\Q{{\bf{Q}}}
\def\p{{\bf{p}}}
\def\x{{\bf{x}}}
\def\X{{\bf{X}}}
\def\Y{{\bf{Y}}}
\def\F{{\bf{F}}}
\def\G{{\bf{G}}}
\def\M{{\bf{M}}}
\def\V{\cal V}
\def\R{{\cal{R}}}
\def\tchi{\tilde{\chi}}
\def\tk{\tilde{\bf{k}}}
\def\tK{\tilde{\bf{K}}}
\def\tq{\tilde{\bf{q}}}
\def\tQ{\tilde{\bf{Q}}}
\def\si{\sigma}
\def\ep{\epsilon}
\def\al{\alpha}
\def\be{\beta}
\def\ep{\epsilon}
\def\up{\uparrow}
\def\de{\delta}
\def\De{\Delta}
\def\dtau{ \Delta \tau} 
\def\up{\uparrow}
\def\dwn{\downarrow}
\def\ksi{\xi}
\def\etha{\eta}
\def\product{\prod}
\def\goto{\rightarrow}
\def\switch{\leftrightarrow}
                           
\title{A Quantum Monte Carlo algorithm for non-local corrections
to the Dynamical Mean-Field Approximation}
\author{M.~Jarrell, Th.~Maier, C.~Huscroft, and S.~Moukouri}
\address{University of Cincinnati, Cincinnati OH 45221, USA}

\maketitle

\begin{abstract}
We present the algorithmic details of the dynamical cluster approximation
(DCA), with a quantum Monte Carlo (QMC) method used to solve the effective
cluster problem.  The DCA is a fully-causal approach which systematically
restores non-local correlations to the dynamical mean field approximation
(DMFA) while preserving the lattice symmetries.  The DCA becomes exact for 
an infinite cluster size, while reducing to the DMFA for a cluster size of 
unity.  We present a generalization of the Hirsch-Fye QMC algorithm for the 
solution of the embedded cluster problem. We use the two-dimensional Hubbard 
model to illustrate the performance of the DCA technique.  At half-filling, 
we show that the DCA drives the spurious finite-temperature antiferromagnetic
transition  found in the DMFA slowly towards zero temperature as the cluster size increases, in conformity with the Mermin-Wagner theorem.
Moreover, we find that there is a finite temperature metal to insulator 
transition which persists into the weak-coupling regime.  This suggests that 
the magnetism of the model is Heisenberg like for all non-zero interactions. 
Away from half-filling, we find that the sign problem that arises in QMC 
simulations is significantly less severe in the context of DCA. Hence, we 
were able to obtain good statistics for small clusters. For these clusters, 
the DCA results show evidence of non-Fermi liquid behavior and 
superconductivity near half-filling.   

\end{abstract}

\section{Introduction}
\label{Sec_INTRO}

One of the most active subfields in condensed matter theory is the 
development of new algorithms to simulate the many-body problem. This 
interest is motivated by various physical phenomena, including high 
temperature superconductivity, magnetism, heavy fermions and the rich 
phenomenology occurring in quasi-one dimensional compounds. In the last 
few years, important progress has been made. Well-controlled results 
have been obtained by exact diagonalization and quantum Monte Carlo 
methods (QMC)\cite{dagotto}.  However, these algorithms suffer from a 
common limitation in that the number of degrees of freedom grows rapidly 
with the lattice size.  As a consequence, the calculations are restricted 
to relatively small systems. In most cases, the limited size of the 
system prohibits the study of the low-energy physics of these models.

Recently, another route to quantum simulations has been proposed. 
Following Metzner and Vollhardt\cite{metzvoll} and 
M\"uller-Hartmann\cite{muller-hartmann} who showed  that in the limit 
of infinite dimensions, the many-body problem becomes purely local, 
a mapping to a self-consistent Anderson impurity problem was 
performed \cite{pruschke,georges}. The availability of many 
techniques to solve the Anderson impurity Hamiltonian has led to a 
dramatic burst of activity.   However, when applied to systems in two 
or three dimensions this self-consistent approximation, referred to 
as the dynamical mean field approximation (DMFA), displays some
limitations. Due to its local nature, the DMFA neglects spatial
fluctuations which are essential when the order parameter is 
non-local, or when short-ranged spin correlations are present. 

	An acceptable theory which systematically incorporates 
non-local corrections to the DMFA is needed. It must be able to 
account for fluctuations in the local environment in a self-consistent 
way, become exact in the limit of large cluster sizes, and recover the 
DMFA when the cluster size equals one.  It must be easily implementable 
numerically and preserve the translational and point-group symmetries 
of the lattice.  Finally, it should be fully 
causal so that the single-particle Green function and self energy are 
analytic in the upper half plane. There have been several attempts to 
formulate theories which satisfy these requirements, but all 
fail in some significant way\cite{agonis}.
	
	In recent publications \cite{DCA_Hettler1,DCA_Hettler2,DCA_Maier1}, 
the dynamical cluster approximation (DCA) has been proposed as an extension 
to the DMFA which satisfies all these criteria. The DCA is built 
in close analogy with the DMFA. In the DCA, the lattice problem is mapped 
to a self-consistently embedded finite-sized cluster, instead of a single 
impurity as in the DMFA.  The key idea of the DCA is to use the irreducible 
quantities (self energy, irreducible vertices) of the embedded cluster as 
an approximation for the corresponding lattice quantities. These irreducible 
quantities are then applied to construct the lattice reducible quantities 
such as the Green function or susceptibilities in the different channels. 
The cluster problem generated by the DCA may be solved by using a variety 
of techniques including the Quantum Monte Carlo (QMC) method \cite{fye}, 
the Fluctuation Exchange (FLEX) approximation \cite{FLEX_Bickers} or the 
Non-Crossing Approximation (NCA)\cite{NCA_Bickers}. 

The QMC method, and the Hirsch-Fye algorithm \cite{fye} in particular, 
is the most reliable of these techniques.  The Hirsch-Fye algorithm was 
originally designed for the treatment of few-impurity problems. Hence, 
it has been widely applied to the Kondo problem \cite{fye} and also to 
solve the impurity problem of the DMFA.  For embedded cluster problems, 
this algorithm shows a mild sign problem, compared to that encountered in 
previous finite-sized simulations, presumably due to the coupling to 
the host.  Thus, we are able to perform simulations at significantly 
lower temperatures than with other available techniques.  However, in 
order to study a meaningful set of clusters of different sizes, it is 
necessary to use massively parallel computers.

Throughout this paper, we will use the two-dimensional Hubbard model 
on a simple square lattice as an example.  Its Hamiltonian is  
\begin{eqnarray}
\label{hubbard}
H&=&-\sum_{\i\j}t_{\i\j}(c_{\i\sigma }^{\dag}c_{\j\sigma }+{\rm{h.c.}})
+\epsilon\sum_{\i\sigma}n_{\i\sigma} \nonumber  \\
&+& U\sum_{\i}(n_{\i\uparrow}-1/2)(n_{\i\downarrow }-1/2) 
\end{eqnarray}
where $t_{\i\j}$ is the matrix of hopping integrals, $c_{\i\sigma }^{(\dag)}$
is the annihilation (creation) operator for electrons on lattice site $\i$ 
with spin $\sigma$, and $U$ the intraatomic repulsion.  We will take $\mu=0$ 
and vary the orbital energy $\epsilon$ to fix the filling.  The model 
has a long history and is still the subject of an intensive research 
in relation with the high-temperature superconductivity, the non-Fermi 
liquid phenomenon, the metal to insulator transition and magnetism in 
various physical systems dominated by strong correlations.  Some short 
accounts\cite{DCA_Huscroft1,DCA_Moukouri1} of the DCA applied to this 
model have been recently published but without a full description 
of the details of the algorithm and numerical subtleties.  It is the 
purpose of this paper to present the full account of the DCA-QMC 
technique.  A typical DCA algorithm using the QMC technique as 
the cluster solver is made of three main blocks: the self-consistent 
loop, the analysis block and the analytical continuation block. The 
self-consistent loop is the most important of the three blocks; it is 
devoted to the mapping of the lattice to the cluster (coarse-graining) 
and to the solution of the cluster problem by the QMC method. In the 
analysis block, cluster Green functions obtained from the QMC method are
transformed to lattice Green functions following the procedure described
in section ~\ref{Sec_DCA}. The last block is devoted to the computation of
the lattice real frequency quantities from the analytical continuation
of the corresponding QMC imaginary-time quantities by the maximum
entropy method (MEM).\cite{JARRELLandGUB}

        This paper is organized as follows.  In the next section,
we review the dynamical mean field approximation.  In Sec.~\ref{Sec_DCA}, 
we review the DCA formalism in which the lattice problem is mapped
to a self-consistently embedded periodic cluster, and discuss the
relationship between the cluster and the lattice. In this section we
describe how different lattice Green functions can be obtained from
the cluster quantities.  In Sec.~\ref{Sec_QMC}, we derive a modified form of
the Hirsch-Fye QMC algorithm, which may be used to solve the effective
cluster problem.  We also discuss the conditioning and optimization of a
variety of one and two-particle measurements.  In Sec.~\ref{DCA_algo},
we discuss the DCA  algorithm.  In Sec.~\ref{Sec_Results}, we will show 
our results for the two-dimensional Hubbard model.  Comparisons between 
the DCA and the results of finite-sized simulations will be made in order 
to outline the complementarity of the two techniques which has been discussed 
in earlier publications \cite{DCA_Huscroft1,DCA_Moukouri1,DCA_Moukouri2}.
At half-filling we discuss the occurrence of  antiferromagnetism and 
the metal to insulator transition.  Away from  half-filling, we show 
the signature of a non-Fermi liquid behavior and  superconductivity for 
small clusters for which the negative sign problem is mild so that good 
statistics can be obtained at low temperatures.  Finally, in 
Sec.~\ref{Sec_Summary}, we draw  the conclusions on the present work and
discuss future applications of the DCA to various physical problems.

\section{The Dynamical mean-field approximation}
\label{Sec_DMFA}

The DCA algorithm is constructed in analogy with the DMFA. The DMFA is 
a local approximation which was used by various authors in perturbative 
calculations as a simplification of the k-summations which render the 
problem intractable\cite{treglia,kuramoto}.  But it was after the work 
of Metzner and Vollhardt \cite{metzvoll} and M\"uller-Hartmann 
\cite{muller-hartmann} who showed that this approximation becomes exact 
in the limit of infinite dimension that it received extensive attention 
during the last decade.  In this approximation, one neglects the spatial 
dependence of the self-energy, retaining only its variation with time. 
Please see the reviews by Pruschke {\it et al}\cite {pruschke} 
and Georges {\it et al}\cite{georges} for a more extensive treatment. 
In this section, we will show that it is possible to re-interpret the DMFA
as a course graining approximation, and then review its derivation.
                                             
\begin{figure}[ht]
\leavevmode\centering\psfig{file=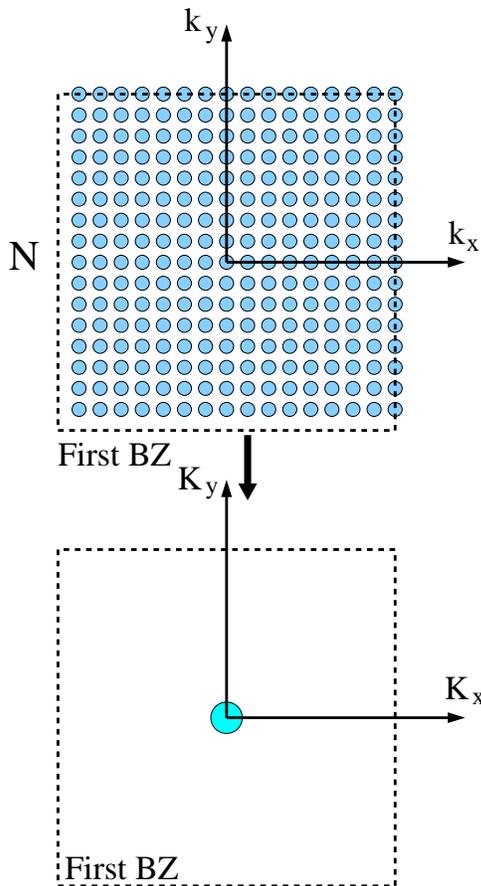,width=2.5 in}
\caption{A single step illustration of coarse-graining in the DMFA: all 
lattice propagators used to calculate the self energy are averaged over 
the points in the first Brillouin zone (top).  This effectively maps
the lattice problem to a single point in reciprocal space (bottom).  
Since the real space and reciprocal space of a single-site cluster are
equivalent, this mapping takes the lattice problem to one of an impurity 
embedded within a host.}
\label{RG_DMFA}
\end{figure}

The DMFA consists of mapping the original lattice problem to a 
self-consistent impurity problem.   As illustrated in  Fig.~\ref{RG_DMFA} 
for a two-dimensional lattice, this is equivalent to averaging the Green 
functions used to calculate the irreducible diagrammatic insertions over 
the Brillouin zone.  An important consequence of this averaging is that 
the self-energy and the irreducible vertices of the lattice are 
independent of the momentum.  Hence, they are those of the impurity.

M\"uller-Hartmann\cite{muller-hartmann} showed that this coarse-graining 
becomes exact in the limit of infinite-dimensions.  For Hubbard-like models, 
the properties of the bare vertex are completely characterized by the Laue 
function $\Delta$ which expresses the momentum conservation at each vertex.  
In a conventional diagrammatic approach
\begin{eqnarray}
\label{Laue_3D}
\Delta(\k_1,\k_2,\k_3,\k_4)=
\sum_\r \exp\left[i\r\cdot(\k_1+\k_2-\k_3-\k_4)\right]\\
=N \delta_{\k_1+\k_2,\k_3+\k_4} 
\nonumber
\end{eqnarray}
where $\k_1$ and $\k_2$ ($\k_3$ and
$\k_4$) are the momenta entering (leaving) each vertex through its
legs of $G$.  However as the dimensionality $D\to\infty$ M\"uller-Hartmann 
showed that the Laue function reduces to\cite{muller-hartmann}
\begin{eqnarray}
\label{Laue_infD}
\Delta_{D\rightarrow\infty}({\bf k}_1,{\bf k}_2,{\bf k}_3,{\bf k}_4)=
1+{\cal O}(1/D)\quad\mbox{.}
\end{eqnarray}
The DMFA assumes the same Laue function, $\Delta_{DMFA}(\k_1,\k_2,\k_3,\k_4)=1$,
even in the context of finite dimensions.  Thus, the conservation of momentum
at internal vertices is neglected. Therefore we may freely sum over the 
internal momentum labels of each Green function leg. This leads to a collapse 
of the momentum dependent contributions and only local terms remain.

This argument may then be applied to the free energy functional.  As 
discussed in many-body texts\cite{agd}, the additional free energy due 
to an interaction may be described by a sum over all closed connected 
graphs.  These graphs may be further separated into compact and 
non-compact graphs.  The compact graphs, which comprise the generating 
functional $\Phi$, consist of the sum over all single-particle 
irreducible graphs.  The remaining graphs, comprise the non-compact 
part of the free energy.  In the infinite-dimensional limit, $\Phi$ 
consists of only local graphs, with non-local corrections of order 
$1/D$.  However, for the non-compact parts of the free energy, non-local 
corrections are of order one, so the local approximation applies only 
to $\Phi$.  Thus, whereas irreducible quantities such as the self energy 
are momentum independent, the corresponding reducible quantities such 
as the lattice Green function are momentum dependent.

The perturbative series for $\Sigma$ in the local approximation is 
identical to that of the corresponding impurity model. However in 
order to avoid overcounting the local self-energy $\Sigma(i\omega_n)$, 
it is necessary to exclude $\Sigma(i\omega_n)$, $i\omega_n=(2n+1)\pi T$ 
is the Matsubara frequency,  from the bare local propagator ${\cal G}$.
${\cal G}(i\omega_n)^{-1}=G(i\omega_n)^{-1} + \Sigma(i\omega_n)$ where 
$G(i\omega_n)$ is the full local Green function. Hence, in the local 
approximation, the Hubbard model has the same diagrammatic expansion
as an Anderson impurity with a bare local propagator 
${\cal G}(i\omega_n;\Sigma)$  which is determined self-consistently.

An algorithm constructed from this approximation is the following: (i) 
An initial guess for $\Sigma(i\omega_n)$ is chosen (usually from 
perturbation theory). (ii) $\Sigma(i\omega_n)$ is used to calculate the 
corresponding local Green function
\begin{eqnarray}
\label{gloc}
G(i\omega_n)=\int d\eta \frac{\rho^0(\eta)}{i\omega_n-\eta-\epsilon-
\Sigma(i\omega_n)}\,,
\end{eqnarray}
where $\rho^0$ is the non-interacting density of states. (iii) Starting 
from $G(i\omega_n)$ and $\Sigma(i\omega_n)$ used in the second step, the 
host Green function 
${\cal G}(i\omega_n)^{-1}=G(i\omega_n)^{-1} + \Sigma(i\omega_n)$
is calculated which serves as bare Green function of the impurity model. 
(iv) starting with ${\cal G}(i\omega_n)$, the local Green function 
$G(i\omega_n)$ is obtained using the Quantum Monte Carlo method (or 
another technique). (v) Using the QMC output for the cluster Green 
function $G(i\omega_n)$ and the host Green function ${\cal G}(i\omega_n)$ 
from the third step, a new $\Sigma(i\omega_n)
={\cal G}(i\omega_n)^{-1}-G(i\omega_n)^{-1}$ is calculated, which is then 
used in step (ii) to reinitialize the process. Steps (ii) - (v) are 
repreated until convergence is reached. In step (iv) the QMC algorithm of 
Hirsch and Fye \cite{fye} may be used to compute the local Green function 
$G(\tau)$ or other physical quantities in imaginary time.  Local dynamical 
quantities are then calculated by analytically continuing the corresponding 
imaginary-time quantities using the Maximum-Entropy Method 
(MEM) \cite{JARRELLandGUB}.

\section{The Dynamical Cluster Approximation}
\label{Sec_DCA}

        In this section, we will review the formalism which leads
to the dynamical cluster approximation.  Here, we first motivate the
fundamental idea of the DCA which is coarse-graining, we then describe
the mapping to an effective cluster problem and discuss the relationship
between the cluster and lattice at the one and two-particle level.

\subsection{Coarse-Graining}

\begin{figure}[ht]
\leavevmode\centering\psfig{file=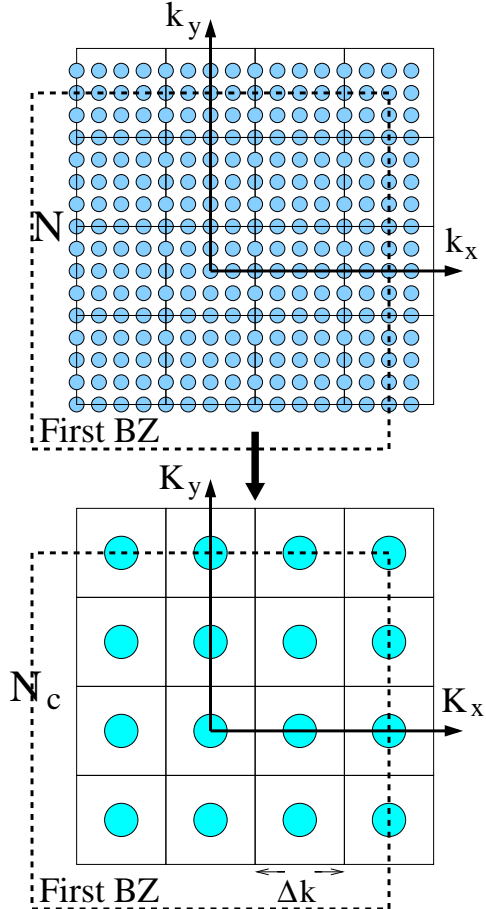,width=2.5 in}
\caption{A single step illustration of coarse-graining in the DCA: all 
lattice propagators used to calculate the self energy are first averaged 
over the points within each cell in the Brillouin zone (top), mapping 
the lattice problem to a small cluster defined by the centers of the 
cells embedded within a host (bottom).}
\label{RG_DCA}
\end{figure}  
Like the DMFA, the DCA may be intuitively motivated with a
coarse-graining transformation. In the DMFA, the propagators
used to calculate $\Phi$ and its derivatives were coarse-grained 
over the entire Brillouin zone, leading to local (momentum
independent) irreducible quantities.  In the DCA, we wish to relax 
this condition, and systematically restore momentum conservation
and non-local corrections.  Thus, in the DCA, the reciprocal space
of the lattice (Fig.~\ref{RG_DCA}) which contains $N$ points 
is divided into $N_c$ cells of identical linear size $\Delta k$. 
The coarse-graining transformation is set by averaging the Green 
function within each cell.  If $N_c=1$ the original lattice problem 
is mapped to an impurity problem (DMFA). If  $N_c$ is larger than 
one, then non-local corrections of length $\approx \pi/\Delta k$ to 
the DMFA are introduced.  Provided that the propagators are 
sufficiently weakly momentum dependent, this is a good approximation. 
If $N_c$ is chosen to be small, the cluster problem can be 
solved using conventional techniques such as QMC, NCA or FLEX. 
This averaging process also establishes a relationship between the 
systems of size $N$ and $N_c$.  A simple and unique choice which 
will be discussed in Sec.~\ref{SSec_DCA_diagram} is to equate the 
irreducible quantities (self energy, irreducible vertices) of the 
cluster to those in the lattice.

\subsection{A diagrammatic derivation}
\label{SSec_DCA_diagram}

        This coarse graining procedure and the relationship of 
the DCA to the DMFA is illustrated by a microscopic diagrammatic 
derivation of the DCA.  
\begin{figure}[ht]
\leavevmode\centering\psfig{file=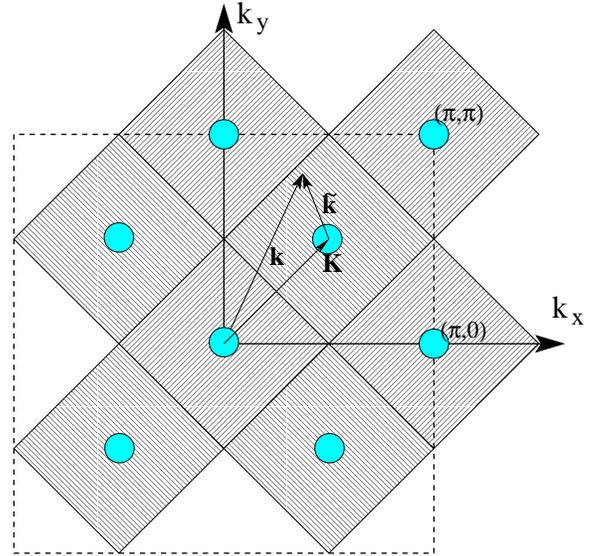,width=3. in}   
\caption{Coarse-graining cells for $N_c=8$ (differentiated by alternating 
fill patterns) that partition the first Brillouin Zone (dashed 
line).  Each cell is centered on a cluster momentum $\K$ (filled 
circles). To construct the DCA cluster, we map a generic momentum 
in the zone such as $\k$ to the nearest cluster point $\K=\M(\k)$ 
so that $\kt=\k-\K$ remains in the cell around $\K$.  }
\label{BZ_Nc8}
\end{figure}
The DCA systematically restores the momentum conservation at internal 
vertices relinquished by the DMFA.  The Brillouin-zone is divided into 
$N_c=L^D$ cells of size $\Delta k=2\pi/L$ (c.f.~Fig.~\ref{BZ_Nc8} for 
$N_c=8$).  Each cell is represented by a cluster momentum $\bf K$ 
in the center of the cell. We require that momentum conservation 
is (partially) observed for momentum transfers between cells, 
i.e., for momentum transfers larger than $\Dk$, but neglected 
for momentum transfers within a cell, i.e., less than $\Dk$. This 
requirement can be established by using the Laue function \cite{DCA_Hettler2}
\begin{equation}
\label{Laue_DCA}
\Delta_{DCA}(\k_1,\k_2,\k_3,\k_4)=
N_c \delta_{\M(\k_1)+\M(\k_2),\M(\k_3)+\M(\k_4)} \quad\mbox{,}
\end{equation}
where $\M(\k)$ is a function which maps $\k$ onto the momentum 
label $\K$ of the cell containing $\k$ (see, Fig.~\ref{BZ_Nc8}).  
This choice for the Laue function systematically interpolates 
between the exact result, Eq.~\ref{Laue_3D}, which it recovers 
when $N_c\to N$ and the DMFA result, Eq.~\ref{Laue_infD}, 
which it recovers when $N_c=1$.  With this choice of the Laue 
function the momenta of each internal leg may be freely summed 
over the cell.

\begin{figure}[ht]
\leavevmode\centering\psfig{file=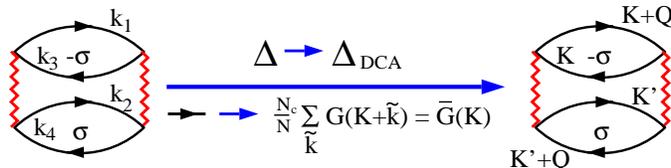,width=3.5 in}   
\caption{A second-order term in the generating functional of the 
Hubbard model.  Here the undulating line represents the interaction
$U$, and on the LHS (RHS) the solid line the lattice (coarse-grained) 
single-particle Green functions.  When the DCA Laue function is used 
to describe momentum conservation at the internal vertices, the 
momenta collapse onto the cluster momenta and each lattice Green 
function is replaced by the coarse-grained result.}
\label{collapse_DCA}
\end{figure}

This is illustrated for the second-order term in the generating 
functional in Fig.~\ref{collapse_DCA}.  Each internal leg $G(\k)$ 
in a diagram is replaced by the coarse--grained Green function 
${\bar G}(\M(\k))$, defined by
\begin{equation}
\label{Gbar}
\bar{G}(\K) \equiv \frac{N_c}{N}\sum_{\kt}G(\K+\kt) \quad\mbox{,} 
\end{equation}
where $N$ is the number of points of the lattice, $N_c$ is the number 
of cluster $\K$ points, and the $\kt$ summation runs over the momenta 
of the cell about the cluster momentum  $\K$ (see, Fig.~\ref{BZ_Nc8}). 
The diagrammatic sequences for the generating functional and its 
derivatives are unchanged; however, the complexity of the problem is 
greatly reduced since $N_c\ll N$.  

As with the DMFA, the coarse-graining approximation will be applied to 
only the compact part of the free energy, $\Phi$, and its derivatives.  
This is justified by the fact that there is no need to coarse-grain
the remaining terms in the free energy.  Formally, we have justified
this elsewhere by exploring the $\Dk$-dependence of the compact and 
non-compact parts of the free energy\cite{DCA_Karan1}.  The generating 
functional is the sum over all of the closed connected compact diagrams, 
such as the one shown in Fig.~\ref{collapse_DCA}.  The corresponding 
DCA estimate for the free energy is 
\begin{equation}
F_{DCA}= -k_B T\left(
\Phic-\mbox{tr}\left[{\bf{\Sigma}}_\sigma {\bf{G}}_\sigma\right] 
-\mbox{tr}\ln\left[-{\bf{G}}_\sigma\right]\right)
\end{equation}
where $\Phic$ is the cluster generating functional.  The trace indicates 
summation over frequency, momentum and spin.  $F_{DCA}$ is stationary 
with respect to ${\bf{G}}_\sigma$,
\begin{equation}
\frac{-1}{k_B T}\frac{\delta F_{DCA}}{\delta G_\sigma(\k)}=
\Sigmac_\sigma(\M(\k))-\Sigma_\sigma(\k)=0,
\end{equation}
which means that $\Sigma(\k)=\Sigmac_\sigma(\M(\k))$ is the proper 
approximation for the lattice self energy corresponding to $\Phic$. 
The corresponding lattice single-particle propagator is then given by 
\begin{equation}
G(\k,z) =\frac{1}{z-\ep_\k-\ep-\Sigmac(\M(\k),z) } \,.
\label{G_DCA}
\end{equation}

        A variety of techniques may be used to sum the cluster diagrams
in order to calculate $\Sigmac$ and the vertex functions $\Gammac$.  In 
the past, we have used QMC\cite{DCA_Huscroft1}, the non-crossing 
approximation\cite{DCA_Maier1} or the Fluctuation-Exchange 
approximation.  Here, we will use the QMC technique which we will 
detail in Sec.~\ref{Sec_QMC}.  Since QMC is systematically exact; i.e.\ 
it effectively sums all diagrams to all orders, care must be taken when 
defining the initial Green function (the solid lines in 
Fig.~\ref{collapse_DCA}) to avoid overcounting diagrams on the 
cluster.  For example, to fourth order and higher in perturbation 
theory for the self energy, non-trivial self energy corrections 
enter in the diagrammatic expansion for the cluster self energy of 
the Hubbard model.  To avoid overcounting these contributions, we must 
first subtract off the self energy corrections on the cluster from the 
Green function line used to calculate $\Sigma_c$ and its functional 
derivatives.  This cluster-excluded Green function is given by
\begin{equation}
\label{Gscript}
\frac{1}{\Gscript(\K,z)} = \frac{1}{\Gbar(\K,z)} + \Sigmac(\K,z)
\end{equation}
which is the coarse-grained Green function with correlations on the
cluster excluded.  Since $\Sigmac(\K,z)$ is not known {\em{a priori}}, 
it must be determined self-consistently, starting from an initial 
guess, usually from perturbation theory.  This guess is used to 
calculate $\Gbar$ from Eq.~\ref{Gbar}.  $\Gscript(\K,z)$ is then 
calculated with Eq.~\ref{Gscript}, and it is used to initialize
the QMC calculation.  The QMC estimate for the cluster self energy 
is then used to calculate a new estimate for $\bar{G}(\K)$ using 
Eq.~\ref{Gbar}.  The corresponding ${\cal{G}}(\K)$ is used to 
reinitialize the procedure which continues until $\Gc=\Gbar$ and 
the self energy converges to the desired accuracy.

One of the difficulties encountered in earlier attempts to include 
non-local corrections to the DMFA was that these methods were not 
causal\cite{peter1,avi}.  The spectral weight was not conserved and 
the imaginary parts of the one-particle retarded Green functions 
and self-energies were not negative definite as required by causality.  
The DCA algorithm presented in this subsection does not present 
these problems. This algorithm is fully causal as shown by Hettler 
{\it et al.}\cite{DCA_Hettler2}. They analyze the different steps of 
the self-consistent loop and found that none of them breaks the 
causality of the Green functions. Starting from the QMC block, one 
can see that if the input  ${\cal {G}}$ is causal, since the QMC 
algorithm is essentially exact, the output  $\Gc$ will also be causal.   
Then the corresponding $\Sigma_c(\K,i\omega_n)$ is causal.  This in turn 
ensures that the coarse-grained Green function $\bar{G}(\K,i\omega_n)$ 
also fulfills causality. The only non-trivial operation which may 
break causality is the calculation of ${\cal {G}}(\K,i\omega_n)$. 
Hettler {\it et al.} used a geometric proof to show that even this 
part of the loop respects causality.   

In the remainder of this section, we will give further details 
about the DCA formalism, and discuss the relationship between 
the cluster and the lattice problems.  Below, we will discuss 
the steps necessary to choose the coarse-graining cells and 
ensure that symmetries of the lattice are preserved.

\subsection{Selecting the Coarse-Graining cells}
\label{SSec_Cells}

As we will see in Sec.~\ref{Sec_QMC} the solution of the cluster 
problem using the quantum Monte Carlo method though is a great 
simplification over the original lattice problem is still a 
formidable task.  The reason is that the self-consistent nature 
of the cluster problem forces us to adopt the Hirsch-Fye 
algorithm. While this algorithm is very efficient for few impurity 
problems, it becomes slow even for a cluster of a 
modest size. Therefore, in order to study the size dependence 
of physical quantities we adopt various cluster tilings of the 
lattice instead of confining ourselves to only the usual square 
tilings $N_c=4,16,36,64...$.

\begin{figure}[ht]
\leavevmode\centering\psfig{file=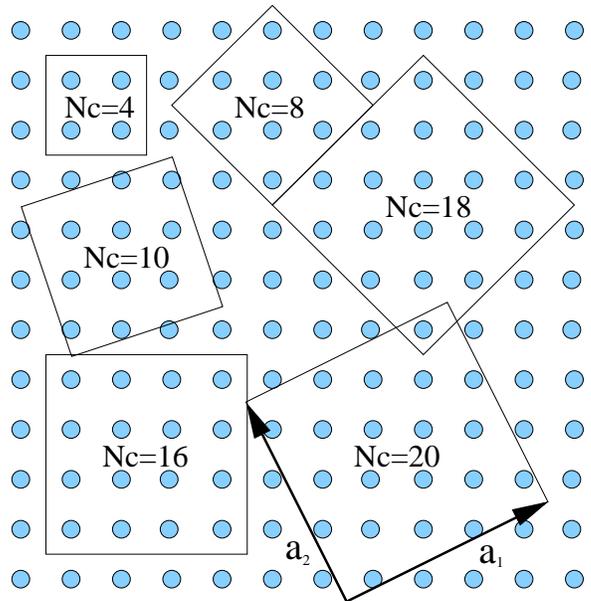,width=3. in}   
\caption{Different tile sizes and orientations in real space.  The 
tiling principle translation vectors, $\a_1$ and $\a_2$, form two 
sides of each tiling square (illustrated for the $N_c=20$ 
tiling).  For square tile geometries, $a_{2x}=-a_{1y}$ and 
$a_{2y}=a_{1x}$.}
\label{tilings}
\end{figure}
\begin{figure}[ht]
\leavevmode\centering\psfig{file=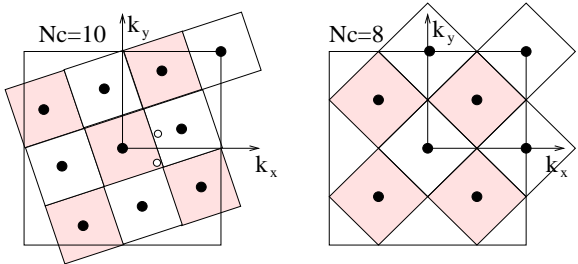,width=3. in}   
\caption{ The coarse graining cells for $N_c=8$ and $10$ each 
centered on a coarse-grained momenta $\K$ represented as black 
filled dots.  For $N_c=8$ equivalent momenta $\k$ are always 
mapped to equivalent coarse-grained momenta $\K$.  However, 
this is not true for $N_c=10$ where, for example, the two 
equivalent momenta shown by open dots are mapped to inequivalent
coarse-grained momenta.}
\label{tilings2}
\end{figure}

When selecting the coarse-graining cells, it is important to 
preserve the point group symmetries of the lattice. For example, 
in this study, we will choose a simple square lattice. Both it 
and its reciprocal lattice share $C_{4v}$ symmetry with eight 
point group operations. We must choose a set of coarse-graining 
cells which preserve the lattice symmetry. This may be done by 
tiling the real lattice with squares, and using the {\bf K}
points that correspond to the reciprocal space of the tiling 
centers.  We also will only consider tilings which contain
an even number of sites, to avoid frustrating the magnetic
correlations on the cluster.  Square tilings with an even number 
of sites include $N_c=4,8,10,16,18,20,26,32,34,36,...$.  The 
first few are illustrated in Fig.~\ref{tilings}. The relation
between the principle lattice vectors of the lattice centers, 
${\bf a}_1$ and ${\bf a}_2$, and the reciprocal lattice takes 
the usual form 
${\bf g}_i=2\pi{\bf a}_i/({\bf a}_1 \times {\bf a}_2)$, with 
${\bf K}_{nm}=n{\bf g}_1+m{\bf g}_2$ for integer n and m. For 
tilings with either $a_{1x}=a_{1y}$ (corresponding to 
$N_c=1,8,18,32...)$ or one of $a_{1x}$ or $a_{1y}$ zero
(corresponding to $N_c=1,4,16,36...)$, the principle reciprocal 
lattice vectors of the coarse-grained system either point along 
the same directions as the principle reciprocal lattice vectors 
of the real system or are rotated from them by $\pi/4$. As a 
result, equivalent momenta {\bf k} are always mapped to equivalent 
coarse-grained momenta {\bf K}. An example for $N_c=8$ is shown in 
Fig.~\ref{tilings2}. However, for $N_c=10,20,26,34...$, the 
principle reciprocal lattice vectors of the coarse-grained system 
do not point along a high symmetry direction of the real lattice. 
Since all points within a coarse-grained cell are mapped to its 
center {\bf K}, this means that these coarse-graining choices 
violate the point group symmetry of the real system. This is 
illustrated for $N_c=10$ in Fig.~\ref{tilings2}, where the two
open dots resting at equivalent points in the real lattice, fall 
in inequivalent coarse-graining cells and so are mapped to 
inequivalent {\bf K} points. Thus the tilings corresponding to 
$N_c=10,20,26,34...$ violate the point-group symmetry of the real 
lattice and will be avoided in this study.

        One should note that the coarse-graining scheme also 
depends strongly on dimensionality.  For example, in one 
dimension, any cell with an even number of sites will preserve 
the lattice symmetry.  
 
\subsection{One-particle Green functions}

In the DMFA, after convergence, the local Green function of the lattice 
is identical to that of the impurity model. Though in the DCA, the 
coarse-grained Green function ${\bar G(\K,i\omega_n)}$ is equal to the 
cluster Green function $G_c(\K,i\omega_n)$, this quantity is not, however, 
used as an approximation to the true lattice Green function $G(\K,i\omega_n)$.  
The correct procedure to calculate the lattice physical quantities 
within the DCA is to approximate the lattice irreducible quantities 
with those of the cluster. The lattice reducible quantities are then 
deduced from the irreducible.  This procedure was justified formally 
in Sec.\ref{SSec_DCA_diagram}.  To obtain a physical understanding, 
one must first understand why reducible and irreducible quantities 
must be treated differently.  
Consider a quasiparticle propagating through the system.  The 
screening cloud is described by the single-particle self-energy 
$\Sigma(\k,\omega)$ which itself may be considered a functional of the 
interaction strength $U$ and the single-particle propagator 
$G(\k,\omega)$, $\Sigma=\Sigma[U, G]$.  The different screening 
processes are described perturbatively by a sum of self-energy 
diagrams.   If the size of the screening cloud $r_s$ is short,  the 
propagators which describe these processes need only be accurate for 
distances $< r_s$.  From the Fourier uncertainty principle, we know that the 
propagators at short distances may be accurately described by a coarse 
sampling of the reciprocal space, with sampling rate $\Delta k= \pi/r_s$. 
Hence, in this case, $\Sigma[U,G]$ may be quite well approximated by 
$\Sigma[U,\bar G]$.

On the other hand, the phase accumulated as the particle propagates 
through the system is described by the Fourier transform of the 
single-particle Green function.  Since this accumulated phase is crucial 
in the description of the quantum dynamics it is important that $G(r)$ 
remains accurate at long distances, so it should not be coarse-grained 
as described above.   However it may be constructed from the approximate 
self-energy.  Hence, the approximate lattice Green function is given by
Eq.~\ref{G_DCA}.  Thus, as in the DMFA, the lattice Green function is 
generally more strongly momentum dependent than the corresponding self 
energy.

In the case of the 2D Hubbard model, non-local correlations are the most 
important in the parameter regime close to the quantum critical point at 
half filling.  Away from this parameter regime $r_s$ is thus expected to 
be short. Here, the above construction scheme for the approximate lattice 
Green function is likely to yield accurate results even for clusters of 
modest size. However, as the quantum critical point is approached, longer 
range correlations are important.  As a consequence one will need to 
evaluate $\Sigma[U,\bar G]$ on larger clusters.

\subsection{Two-Particle Green Functions}
\label{SSec_twopart}

A similar procedure is used to construct the two-particle quantities
needed to determine the phase diagram or the nature of the dominant
fluctuations that can eventually destroy the quasi-particle.  This 
procedure is a generalization of the method of calculating response 
functions in the DMFA\cite{zlatic,DMFA_Jarrell1}. In the DCA, 
the introduction of the momentum dependence in the self-energy will 
allow one to detect some precursor effects which are absent in the DMFA;
but for the actual determination of the nature of the instability,
one needs to compute the response functions. These susceptibilities 
are thermodynamically defined as second derivatives of the free energy 
with respect to external fields.  $\Phic({\bf{G}})$ and $\Sigmac_\sigma$, 
and hence $F_{DCA}$ depend on these fields only through $G_\sigma$ and 
$G_\sigma^0$.  Following Baym\cite{baym} it is easy to verify that, the 
approximation 
\begin{equation}
\Gamma_{\sigma,\sigma'}\approx \Gammac_{\sigma,\sigma'} 
\equiv \delta\Sigmac_\sigma/\delta G_{\sigma'}
\end{equation}
yields the same estimate that would be obtained from the second derivative 
of $F_{DCA}$ with respect to the applied field.  For example, the first 
derivative of the free energy with respect to a spatially 
homogeneous external magnetic field $h$ is the magnetization,
\begin{equation}
m=\mbox{tr}\left[\sigma {\bf{G}}_\sigma\right].
\end{equation}
The susceptibility is given by the second derivative, 
\begin{equation}
\frac{\partial m}{\partial h}=
\mbox{tr}\left[\sigma\frac{\partial {\bf{G}}_\sigma}{\partial h}\right].
\end{equation}
We substitute ${\bf{G}}_\sigma =
\left( {\bf{G}}_\sigma^{0-1}- {{\bf{\Sigmac}}}_\sigma\right)^{-1}$, and
evaluate the derivative, 
\begin{equation}
\frac{\partial m}{\partial h}=
\mbox{tr}\left[\sigma\frac{\partial {\bf{G}}_\sigma}{\partial h}\right]
=\mbox{tr}\left[
{\bf{G}}_\sigma^2
\left( 1 +
\sigma\frac{\partial {{\bf{\Sigmac}}}_\sigma}{\partial {\bf{G}}_{\sigma'}}
  \frac{\partial {\bf{G}}_{\sigma'}}{\partial h}
\right)
\right].
\end{equation}
If we identify 
$\chi_{\sigma,\sigma'}=\sigma \frac{\partial{\bf{G}}_{\sigma'}}{\partial h}$,
and $\chi_{\sigma}^0= {\bf{G}}_\sigma^2$, collect all of the terms 
within both traces, and sum over the cell momenta $\tk$, we obtain the 
two--particle Dyson's equation 
\begin{eqnarray} 
2\big({\bar{\chi}}_{\sigma,\sigma}&-&{\bar{\chi}}_{\sigma,-\sigma}\big)\\
&=&
2{\bar{\chi}}_{\sigma}^0 +
2{\bar{\chi}}_{\sigma}^0 \left( {{\bf{\Gammac}}}_{\sigma,\sigma} 
-{{\bf{\Gammac}}}_{\sigma,-\sigma}\right)
\left({\bar{\chi}}_{\sigma,\sigma}-{\bar{\chi}}_{\sigma,-\sigma} \right)\,.
\nonumber
\end{eqnarray}
We see that again it is the irreducible quantity, i.e. the
vertex function, for which cluster and lattice correspond.

\subsubsection{particle-hole}

In this subsection we will provide more details about the relationship
between the lattice and cluster two-particle Green functions and 
describe how a particle-hole susceptibility may be calculated 
efficiently.  As a specific example, we will describe the calculation 
of the two-particle Green function
\begin{eqnarray}
\label{chiph}
\nonumber
\chi_{\si,\si'}(q,k,k')&=&
\int_{0}^{\beta}\int_{0}^{\beta}\int_{0}^{\beta}\int_{0}^{\beta} 
d\tau_1 d\tau_2 d\tau_3 d\tau_4   \\ 
& \times & e^{i\left( (\omega_{n}+\nu) \tau_1 - \omega_{n} \tau_2 
                +\omega_{n'} \tau_3 - (\omega_{n'}+\nu) \tau_4 \right)} \nonumber \\ 
& \times & 
\langle T_\tau
c_{\k+\q \si}^{\dag}(\tau_1)
c_{\k \si}(\tau_2)
c_{\k' \si'}^{\dag}(\tau_3)
c_{\k'+\q \si'}(\tau_4) 
\rangle  \quad\mbox{,}
\nonumber
\end{eqnarray}             
where  we adopt the conventional notation \cite{agd} $k=(\k,i\omega_n)$,
$k'=(\k,\omega'_n)$, $q=(\q,\nu_n)$ and $T_\tau$ is the time ordering
operator. 

        $\chi_{\si,\si'}(q,k,k')$ and $\Gamma_{\si,\si'}(q,k,k')$ 
are related to each other through
the Bethe-Salpeter equation:
\begin{eqnarray} 
\label{betheph}
\nonumber 
\chi_{\si,\si'}(q,k,k')&=& 
\chi^0_{\si,\si'}(q,k,k') + \chi^0_{\si,\si''}(q,k,k'') \\  
&\times & \Gamma_{\si'',\si'''}(q,k'',k''') \chi_{\si''',\si'}(q,k''',k')
\end{eqnarray} 
where $ \Gamma_{\si,\si'}(q,k,k')$ is the two-particle irreducible vertex 
which is the analogue of the self-energy, $\chi^0_{\si,\si'}(q,k,k'')$ is the 
non-interacting susceptibility constructed from a pair of fully-dressed 
single-particle Green functions. As usual, a summation is to be made for 
repeated indices. 
%In the DMFA,  there is no momentum dependence on the vertex as for the 
%self-energy. $\Gamma[(\q,\nu_n);(\k,i\omega_n);(\k',\omega'_n)]$
%is collapsed to  $\Gamma[\nu_n;i\omega_n;\omega'_n]$ so that the momentum
%dependence of the lattice susceptibilities comes from the non-interacting
%susceptibility alone.  

        We now make the DCA substitution 
$\Gamma_{\si,\si'}(\q,\k,\k') \to \Gammac_{\si,\si'}\left(\q,\M(\k),\M(\k')\right)$
in Eq.~\ref{betheph} (where frequency labels have been suppressed).
Note that only the bare and dressed two-particle Green functions
$\chi$ depend upon the momenta $\kt$ within a cell. Since $\chi$ and 
$\chi^0$ in the product on the RHS of Eq.~\ref{betheph} share no 
common momentum labels, we may freely sum over the  momenta 
$\kt$ within a cell, yielding
\begin{eqnarray} 
\label{betheph_CG}
\nonumber 
\lefteqn{ \chibar_{\si,\si'}(q,K,K')= 
\chibar^0_{\si,\si'}(q,K,K') + \chibar^0_{\si,\si''}(q,K,K'')} \nonumber\\  
&\times & \Gammac_{\si'',\si'''}(q,K'',K''') 
\chibar_{\si''',\si'}(q,K''',K')\,.
\end{eqnarray} 
By coarse-graining the Bethe-Salpeter equation, we have greatly reduced 
its complexity; each of the matrices above is sufficiently small 
that they may be easily manipulated using standard techniques.

In contrast with the single-particle case where the coarse-grained
quantities are identical to those of the cluster, $\chic_{\si,\si'}(q,K,K')$ 
is not equal to  $\chibar_{\si,\si'}(q,K,K')$.  This is because the 
self-consistency is made only at the single-particle level.  Unlike the 
single particle case where both  $\Sigma(K)$ and ${\bar G}(K)$  are 
directly calculated, neither $\Gamma_{\si,\si'}(q,K,K')$ nor the 
coarse-grained susceptibility  $\chibar_{\si,\si'}(q,K,K')$ are 
calculated during the self-consistency.  Instead, the coarse-grained 
non-interacting susceptibility $ \chibar^0_{\si,\si'}(q,K,K')$ is 
calculated in a separate program after the DCA converges using the 
following relation
\begin{eqnarray}
\label{chinot}
\nonumber
\chibar^0_{\si,\si'}[(\q,i\nu_n);(\K,i\omega_n);(\K',i\omega'_n)] =
\delta_{\si,\si'}\delta_{ \K,\K'} \delta_{\omega_n,\omega'_n} \\
\times \frac{N_c}{N} 
\sum_{\tk}G_{\si}( \K+{\bf \tk},i\omega_n)
G_{\si}({ \K+{\bf \tk}+\q}, i\omega_n+\nu_n)
  \quad\mbox{.}
\end{eqnarray}
The corresponding cluster susceptibility is calculated in the QMC
process, as discussed in Sec.~\ref{SSec_QMCMEAS} and the vertex function is 
extracted by inverting the cluster two-particle Bethe-Salpeter equation
\begin{eqnarray} 
\label{betheph_C}
\nonumber 
\chic_{\si,\si'}(q,K,K')= \chic^0_{\si,\si'}(q,K,K') + 
\chic^0_{\si,\si''}(q,K,K'') \\  
          \times  \Gammac_{\si'',\si'''}(q,K'',K''') 
                     \chic_{\si''',\si'}(q,K''',K')\,.
\end{eqnarray} 
If we combine Eqs.~\ref{betheph_C} and \ref{betheph_CG}, then the 
coarse-grained susceptibility may be obtained after elimination of 
$\Gamma(q,K,K')$ between the two equations.
It reads
\begin{equation}
\label{chicoars2}
\nonumber
{\bar \chi}^{-1} = \chi_c^{-1} -\chi_c^{0^{-1}} +  {\bar \chi}^{0^{-1}}\,,
\end{equation}      
where, for example, ${\bar \chi}$ is the matrix formed from  
$\chibar_{\si,\si'}(q,K,K')$ for fixed $q$.  The charge $(ch)$ and 
spin $(sp)$ susceptibilities ${\chi}_{ch,sp}(q,T)$
are deduced from $\chibar$
\begin{eqnarray}
\label{chichsp}
{\chi}_{ch,sp}(q,T)=\frac {(k_BT)^2}{N_c^2}
\sum_{ KK'\sigma\sigma'}
\lambda_{\sigma \sigma'} 
\chibar_{\si,\si'}(q,K,K')
  \quad\mbox{,} 
\end{eqnarray}
where $\lambda_{\sigma \sigma'}=1$ for the charge channel and 
$\lambda_{\sigma \sigma'}=\sigma\sigma'$ for the spin channel.

\subsubsection{particle-particle} 
\label{SSSec_particle_particle}
The calculation of susceptibilities in the particle-particle
channel is essentially identical to the above.  The exception 
to this rule occurs when we calculate susceptibilities for 
transitions to states of lower symmetry than the lattice symmetry. 
For example, in order to obtain the pair function of the desired 
symmetry ($s,p,d$), the two-particle Green function must be 
multiplied by the corresponding form factors $g(\k)$ and $g(\k')$.  
In the study of the Hubbard model below, we will be particularly 
interested in $g(\k)=1$ ($s$ wave), $g(\k)=cos (k_x)+cos (k_y)$ 
(extended $s$ wave) and $g(\k)=cos (k_x)-cos (k_y)$ ($d_{x^2-y^2}$ 
wave). These symmetries have been evoked as possible candidates for 
the superconducting ground state.

These factors modify the Bethe-Salpeter equations 
\begin{eqnarray}
\label{bethepp}
\lefteqn{ g(\k)\chi(q,k,k')g(\k') = g(\k)\chi^0(q,k,k')g(\k')} \\
&+& g(\k)\chi^0(q,k,k'')  \times \Gamma(q,k'',k''') 
                          \times \chi(q,k''',k')g(\k')
\nonumber
\end{eqnarray}  
where
\begin{eqnarray}
\label{chipp}
\chi(q,k,k')&=&
\int_{0}^{\beta}\int_{0}^{\beta}\int_{0}^{\beta}\int_{0}^{\beta} 
d\tau_1 d\tau_2 d\tau_3 d\tau_4  \\ 
& \times & 
e^{i\left(
(\omega_{n}+\nu) \tau_1
-\omega_{n} \tau_2 
+\omega_{n'} \tau_3 
-(\omega_{n'}+\nu) \tau_4 
\right)
} \nonumber \\ 
& \times & 
\langle T_\tau
c_{\k+\q  \si}^{\dag}(\tau_1)
c_{-\k   -\si}^{\dag}(\tau_2) 
c_{-\k'  -\si}(\tau_3)
c_{\k'+\q \si}(\tau_4)
\rangle  \quad\mbox{,} \nonumber
\end{eqnarray}        
% Notational convention:
%                   si
%  k+q     --->------------>--- k'+q
% iwn+inu    1   |      |  4    iwn'+inu
%                |      |
%                |      |
%                |      |
%  -iwn      2   | -si  |  3    -iwn'
%  -k      --->------------>--- -k'
%in Eq.~\ref{bethepp}, we describe the evolution of the particle-particle
%pair created in the initial state (12) and destroyed in the final 
%state (34). 
On the LHS, we have dropped the spin indices since we will consider only 
opposite-spin pairing. 
Eq.~\ref{bethepp} cannot be easily solved if it is coarse-grained, since 
this will partially convolve  $\chi(q,k,k')$ with {\em{two}} factors of
$g$ on the LHS and {\em{one}} factor on the RHS.  Hence for the pairing 
susceptibilities, or for any situation where non-trivial form factors 
must be used, we use the equivalent equation involving the reducible 
vertex $T_2$ (instead of the irreducible vertex $\Gamma$)
\begin{eqnarray}
\label{betheppT2}
\nonumber
 g(\k)\chi(q,k,k')g(\k') &=& g(\k)\chi^0(q,k,k')g(\k') \\
                        &+& g(\k)\chi^0(q,k,k'') \nonumber\\ 
                        &\times& T_2(q,k'',k''') \chi^0(q,k''',k')g(\k')\,,
\end{eqnarray}  
where
\begin{eqnarray}
\label{T2}
T_2(q,k,k') &=& \Gamma(q,k,k') \\
&+&\chi^0(q,k,k'')\Gamma(q,k'',k''')\chi^0(q,k''',k') + \cdots \nonumber 
\end{eqnarray}
We define
\begin{eqnarray}
\label{chipart}
\Pi_{g,g}(q,k,k')&=&g(\k)\chi(q,k,k') g(\k')\\
\Pi_{g,g}^0(q,k,k')&=&g(\k)\chi^0(q,k,k') g(\k')\\
\Pi_{g}^0(q,k,k')&=&g(\k)\chi^0(q,k,k')\,.
\end{eqnarray}
The remaining steps of the calculation are similar to the particle-hole
case.  We invert the cluster particle-particle Bethe-Salpeter equation 
with $g=1$ for the cluster, in order to extract $\Gammac$.  We then
coarse-grain Eq.~\ref{T2}, and use $\Gammac$ to calculate the coarse-grained 
${\bar{T_2}} = \Gammac\left( 1-\chibar^0\Gammac\right )^{-1}$.
We then coarse-grain Eq.~\ref{betheppT2}, and use the coarse-grained 
${\bar{T_2}}$ to calculate the coarse-grained 
${\bar{\Pi}}_{g,g}$
\begin{eqnarray}
{\bar\Pi}_{g,g}(q,K,K') &=& {\bar\Pi}_{g,g}^0(q,K,K')  
\label{pair_dyson} \\
&+&{\bar\Pi}_{g}^0(q,K,K'') {\bar{T}}_2(q,K'',K'''){\bar\Pi}_{g}^0(q,K''',K')\,.
\nonumber
\end{eqnarray}
The pairing susceptibility of a desired symmetry is given by
\begin{eqnarray}
\label{chichsp1}
P_g(q,T)=\frac {(k_BT)^2}{N_c^2} \sum_{K,K'} {\bar \Pi}_{gg}(q,K,K')
  \quad\mbox{.}
\end{eqnarray}

\subsection{Local quantities}

We will also need to evaluate a number of local quantities on the lattice.  
They include the magnetic moment, the local magnetic susceptibility,
the local Green function, etc.  The local cluster quantities are identical 
to the local lattice ones.  This may be seen for example on the one-particle 
Green function. The coarse-grained Green function is related to the lattice 
Green function as follows
\begin{eqnarray}
{\bar G}(\r, \omega)=\frac{1}{N} \sum_{\K,\tk}\sum_{\X,\r'}
 e^{i \K \cdot (\r-\r')}e^{i  \tk \cdot (\X+\r')}G(\X+\r',\omega).
\label{local}
\end{eqnarray}
It is easy to see from this relation that ${\bar G}(0,\omega)=
G(0,\omega)$.

\section{The Quantum Monte Carlo algorithm}
\label{Sec_QMC}

        In this section we will derive a generalization of the 
Hirsch-Fye Anderson impurity algorithm suitable to simulate a Hubbard
cluster embedded in a self-consistently determined host.  We will then
discuss the differences between this algorithm and the more familiar
Blanckenbecler-Sugar-Scalapino (BSS) algorithm\cite{BSS} used to simulate
finite-sized systems.  Finally, we will discuss how different quantities
mentioned above may be measured efficiently and how the code can
be optimized.

\subsection{Formalism}
\label{Sec_QMC_Formalism}

The Hirsch-Fye algorithm is an action-based technique.  Therefore, knowledge 
of the underlying Hamiltonian is not required provided that we know the Green 
function for the non-interacting cluster coupled to the host, and the 
interacting part of the action or Hamiltonian.  The interacting part is 
unchanged by the coarse-graining since it is purely local. It may be written 
in the real space as follows, 
\begin{eqnarray}
H_{I} = U \sum_{\i=1}^{N_c}(n_{\i,\up}-1/2)(n_{\i,\dwn}-1/2)\,,
\end{eqnarray}
and the bare cluster Green function is $\Gscript(\K,i\omega_n)$.

Given this information, the most direct way to derive this algorithm 
is to express the partition function as path integrals over Grassmann 
variables.  The first step is to disentangle the interacting, $H_I$, and 
non-interacting, $H_0$, parts of the Hamiltonian using a Trotter-Suzuki 
decomposition for the partition function.   We divide the interval 
$[0,\beta]$ into $N_l$ sufficiently small subintervals $\Delta \tau=\beta/N_l$ 
such that 
$\Delta\tau^2\left[H_0,H_I\right]$ may be neglected. This leads to
\begin{eqnarray}
Z=Tre^{-\be H}=Tr\prod_{l=1}^{N_l}e^{-\De\tau H}
        \approx Tr\prod_{l=1}^{N_l}e^{-\De\tau H_{0}}e^{-\De\tau H_{I}}\,.
\label{Z1}
\end{eqnarray}
The interacting part of the Hamiltonian may be further decoupled by 
mapping it to an auxiliary Ising field via a discrete 
Hirsch-Hubbard-Stratonovich (HHS)\cite{HHS} transformation,                     \begin{eqnarray}
\nonumber
e^{-\De\tau H_{I}} = 
e^{-\De\tau U\sum_{\i}(n_{\i \up}-1/2)(n_{\i \dwn}-1/2) }\\
= \frac{1}{2} e^{-\De\tau U/4} 
\prod_{\i} \sum_{s_{\bf i}=\pm 1} e^{\al s_{\bf i} (n_{\i \up} - n_{\i \dwn})}\,,
\end{eqnarray}
where $\cosh(\al)=e^{\De\tau U/2}$.

We now introduce coherent states of the operators on the cluster 
and in the host as the basis states and express the partition 
function as path integrals over the corresponding Grassmann 
variables $\gamma_{\i,l,\sigma}$ and $\phi_{\k,l,\sigma}$ defined 
over each $N_l$ time slices $\tau_l=l \Delta \tau$ of the interval 
$[0,\beta]$ \cite{NegeleOrland}.  After substituting the Grassmann 
variables one obtains the following approximation for the partition 
function which becomes exact as $\De\tau \rightarrow 0$,
\begin{eqnarray}
Z \approx \int {\cal D}[\gamma]{\cal D}[\phi]
 e^{-S_0[\gamma,\phi]}e^{-S_I[\gamma]},
\label{partition}
\end{eqnarray}
where $\cal D[...]$ symbols denote the measures of path integration
over the corresponding Grassmann fields and $S_{(0)I}$ is the 
(non-)interacting part of the action.
The interacting part of the action, becomes
\begin{eqnarray}
S_I[\gamma]=- \sum_{\i=1}^{N_c}  \sum_{l=1}^{N_l} \sum_\sigma 
\alpha  \gamma_{\i,l,\sigma}^{*} \sigma s_{\i,l} \gamma_{\i,l-1,\sigma}.
\label{SI}
\end{eqnarray}
The non-interacting parts are
\begin{eqnarray}
\nonumber
S_0[\gamma,\phi] &=& \dtau \sum_{\k,l} \bigg[
\phi_{\k,l,\sigma}^*\left(\frac{\phi_{\k,l,\sigma}-\phi_{\k,l-1,\sigma}}{\dtau} \right) \\
& & +H_{host}(\phi^*_{\k,l,\sigma},\phi_{\k,l-1,\sigma}) \bigg ] \nonumber \\
&+& \dtau \sum_{\i,l,\sigma}
\gamma_{\i,l,\sigma}^*\left(\frac{\gamma_{\i,l,\sigma}-\gamma_{\i,l-1,\sigma}}{\dtau} \right)
\label{S0} \\
&+&\dtau \sum_{\k,\i,l,\sigma}H_{cluster}^0(\phi^*_{\k,l,\sigma},\phi_{\k,l-1,\sigma};\gamma^*_{\i,l,\sigma},\gamma_{\i,l-1,\sigma}) \nonumber
\end{eqnarray}
where $H_{host}$, $H_{cluster}^0$ are the Hamiltonian for the host, and 
the non-interacting degrees of freedom on the cluster including the 
coupling to the host, respectively.  
The detailed form of both $H_{host}$ and $H_{cluster}^0$ are unknown, 
due to the self-consistent renormalization of the host.  However, both
are purely bilinear, and may be integrated out of the action
{\em{without further approximation}}.  

	We will first integrate out the host degrees of freedom.
The partition function becomes
\begin{equation}
Z \propto  
\left[ \product_{\k,\sigma} \det \left({g_{\k,\sigma}} \right)^{-1} \right] 
\int {\cal D}[\gamma]{\cal D}[\phi] e^{-S_c[\gamma]} 
\label{Z2}
\end{equation}
where ${g_{\k,\sigma}}$ is the Green function of the host.  It remains fixed 
during the QMC process, and it may be disregarded since, as we show below, 
we only require knowledge of the ratio of the partition functions for two 
different configurations of the HHS fields.  Other fixed prefactors 
(depending upon $U$, $\beta$ $\cdots$) have also been disregarded in 
Eq.~\ref{Z2}.  $S_c$ is the cluster action.  It takes the form
\begin{equation}
S_c[\gamma] =  \sum_{\i,l;\i',l',\sigma}
\gamma_{\i,l,\sigma}^* \Gscript^{-1}(\i,l;\i',l')\gamma_{\i',l',\sigma} + 
S_I[\gamma]
\label{SC}
\end{equation}
where $\Gscript(\i,l;\i',l')$ is the cluster excluded (i.e.\ non-interacting
on the cluster) Green function, defined previously.  Now we will integrate
out the remaining cluster Grassmann variables.  The partition function
then becomes
\begin{equation}
Z\propto 
Tr_{\{s_\i,l\}} \product_{\sigma} 
\det \left(\Gc_{\sigma; s_{\i,l}} \right)^{-1} 
\label{Z3}
\end{equation}
where again factors which are fixed during the QMC process have been
ignored.  $\left( \Gc_{\sigma; s_{\i l}} \right)^{-1}$ is the inverse 
cluster Green function matrix with elements
\begin{equation}
\left(\Gc_{\sigma; s_{\i l}}\right)^{-1}_{\i,\j,l,l'}
= \delta_{\i,\j} \delta_{l',l-1}
\alpha \sigma s_\i(\tau_l)
+ \Gscript^{-1}_{\i,\j,l,l'}\,.
\label{ithG}
\end{equation}

 	If we re-exponentiate the first term in the RHS of the above 
formula by defining ${\V}_\si(\i,l) \equiv \al s_{i,l}\si$, we can 
write Eq.~\ref{ithG} in a simple matrix notation as 
\begin{eqnarray}
\Gc_{\si}^{-1} = \Gscript^{-1}+ T \left( e^{{\V}_{\si}}-1\right) \;,
\label{Grexp}
\end{eqnarray}
where $T$ is $\de_{i,j}\de_{l-1,l'}$.  The matrix product 
$\Gc_{\si}^{-1} e^{-{\V}_{\si}}$
depends upon the HHS fields only along its diagonal elements.
As can be seen from Eq.~\ref{S0}, each diagonal element of the matrices 
$\Gscript^{-1}$ and hence $\Gc_\sigma^{-1}$ is $1$.  Therefore, the inverse Green 
functions for two different field configurations, $\{s_{\i l}\}$ and 
$\{s_{\i l}'\}$, are related by
\begin{equation}
\Gc_{\si}^{'-1} e^{-{\V'}_{\si}}=\Gc_{\si}^{-1} e^{-{\V}_{\si}}
- e^{-{\V}_{\si}}
+ e^{-{\V'}_{\si}} \,.
\end{equation}
Or, after multiplying by $e^{{\V'}_{\si}}$, and collecting terms
\begin{eqnarray}
         {\Gc'}_{\si}^{-1}-\Gc_{\si}^{-1} =
         (\Gc_{\si}^{-1}-1)e^{-{\V}_{\si}}(e^{{\V}'_{\si}}-e^{{\V}_{\si}}) \,.
\end{eqnarray}
Multiplying from the left by $\Gc$ and from the right by ${\Gc'}$, we find
\begin{equation}
{\Gc'}_{\si} =\Gc_{\si}+
(\Gc_{\si}-1)(e^{ {\V}'_{\si} - {\V}_{\si}}-1){\Gc'}_{\si}
\label{GGprime1}
\end{equation}
or
\begin{equation}
G_{\si}{\Gc'}_{\si}^{-1}=1+(1-\Gc_{\si})(e^{  {\V}'_{\si} - {\V}_{\si} }-1) \,.
\label{GGprime2}
\end{equation}

\subsection{The QMC algorithm}

\label{SSec_QMC_ALG}

We will now proceed to derive the Monte Carlo algorithm. The QMC 
algorithm involves changes in the Hubbard-Stratonovich field 
configuration ${\{s_{\i,l}\}} \to {\{{s'}_{\i,l}\}}$, and accepts 
these changes with the transition probability $P_{s\to s'}$.  Thus, 
to define the algorithm, we need $P_{s\to s'}$ and a relation 
between the cluster Green functions $G$ and $G'$ for the two different 
auxiliary field configurations.  To simplify the notation, we 
introduce a combined space-time index $i=(\i,l)$, and will
consider only local changes in the fields $s_m\to {s'}_m=-s_m$. 
As can be inferred from Eq.~\ref{Z3}, the probability of a 
configuration $\{s_i\}$ is 
$P_{s} \propto \det(\Gc_{\up\{s_i\}}^{-1})\det(\Gc_{\dwn\{s_i\}}^{-1})$;
on the other hand detailed balance requires 
$P_{s'}P_{s'\to s}=P_{s}P_{s\to s'}$ for all $s'$.  We may satisfy 
this requirement either by defining the transition probability 
$P_{s' \to s}=R/(1+R)$, where 
\begin{equation}
R\equiv \frac{P_{s}}{P_{s'}}=
\frac{ \det({{\Gc}'}_{\up})\det({{\Gc}'}_{\dwn}) }
{ \det(\Gc_{\up})\det(\Gc_{\dwn}) } 
\end{equation}
is the relative weight of two configurations, or by letting
$P_{s' \to s}= \mbox{minimum}(R,1)$ (the first choice is called the 
``heat bath'' algorithm, and the second the ``Metropolis'' algorithm).  
If the difference between two configuration is due to a flip of
a single Hubbard Stratonovich field at the $m$th location in the
cluster space-time\cite{fye}, then from Eq.~\ref{GGprime2}
\begin{eqnarray}
R = \prod_{\sigma}[1 + (1-\Gc_{\sigma m,m})(e^{-\al\si (s_m-s_m')} -1)] ^{-1}\,.
\end{eqnarray}
For either the Metropolis or the heat bath algorithm, if the change is 
accepted, then we must update the Green function accordingly. The 
relationship between $G$ and $G'$ is defined by Eq.~\ref{GGprime1} 
\begin{eqnarray}
{{\Gc}'}_{\sigma i j}&=&\Gc_{\sigma i j}\nonumber \nonumber \\
                  &+&\frac{(\Gc_{\sigma i m}-\de_{i,m}) 
         (e^{-\alpha\sigma (s_m-{s'}_m)}-1)}{1+(1-\Gc_{\sigma m,m})
                (e^{-\alpha\sigma (s_m-{s'}_m)}-1)}\Gc_{\sigma m j}\,.
\label{updateG}
\end{eqnarray}

The QMC procedure is initialized by setting 
$\Gc_{\sigma i j }={\cal G}_{i j}$ where ${\cal G}_{i j}$ is the (cluster)
Fourier transform of ${\cal{G}}(\K)$ (Eq.~\ref{Gscript}), and choosing the 
corresponding field configuration with all $s_i=0$.  Then we use 
Eq.~\ref{updateG} to create a Green function corresponding to a meaningful 
field configuration (i.e.\ $s_i=\pm 1$, for each $i=(\i,l)$ or the 
$\{s_i\}$ from a previous run or iteration).  We proceed by sequentially 
stepping through the space-time of the cluster, proposing local changes 
$s_i\to-s_i$.  We accept the change if $P_{s'\to s}$ is greater than a 
random number between zero and one and update the Green function according 
to Eq.~(\ref{updateG}).  After twenty to one hundred warm-up sweeps 
through the space-time lattice of the cluster, the system generally comes 
into equilibrium and we begin to make measurements.  A few lattice 
updates are used between each measurement step to reduce the correlations 
between measurements. This improves the efficiency of the algorithm, since 
as we will see below, the measurements are numerically expensive.  After 
many iterations of lattice updates, numerical round-off error begins to 
accumulate in the Green function update, Eq.~\ref{updateG}.  To compensate 
for this round-off error, the Green functions must be refreshed by again 
setting $\Gc_{\sigma i j }={\cal G}_{i j}$, and then using Eq.~\ref{updateG} 
to recalculate the Green function corresponding to the present field 
configuration.  

\subsection{Differences with the BSS Algorithm}
\label{HF_vs_BSS}

	The Hirsch-Fye (HF) algorithm differs in several ways from the 
more familiar Blanckenbecler-Sugar-Scalapino (BSS) algorithm\cite{BSS} 
used to simulate finite-sized systems.  

The BSS algorithm is more efficient.  HF simulations can be computationally 
quite expensive since the memory and the CPU time required by 
this algorithm scale as $(N_c N_{l})^2$ and $(N_c N_l)^3$, where $N_c$ and 
$N_l$ are respectively the number of cluster sites and the number of time 
slices.  The BSS algorithm scales as $N_l N_{c}^2$ for the memory and 
$N_l N_{c}^3$ for the CPU time.  In order to study a meaningful set of 
cluster sizes using the Hirsch-Fye algorithm, it is necessary to use 
massively parallel computers. The maximum size we studied is $N_c=64$ 
for the two-dimensional Hubbard model. This maximum size is indeed 
smaller than what can be reached with the BSS algorithm applied for 
finite system simulations (FSS). But, one should bear in mind that, in 
the DCA, the system is in the thermodynamic limit, it is the range of 
spatial correlations which is restricted to the cluster size. Cluster 
size effects are of different nature than that occurring in FSS.  
Therefore, the DCA as discussed in previous studies, can provide
information which cannot be obtained from the FSS.

	 The Hirsch-Fye algorithm is action-based, whereas the BSS 
algorithm is Hamiltonian based.  Therefore, the BSS algorithm cannot be 
employed to solve the DCA cluster problem, since the cluster problem 
has no Hamiltonian 
formulation with known parameters, and its action is highly non-local 
in time.  The BSS algorithm requires that the action be local in time.  
The cluster action, Eq.~\ref{SC}, is long-ranged in time due to the 
term involving $\Gscript$.  Thus, the Hirsch-Fye algorithm is the most 
appropriate QMC algorithm to solve the DCA embedded cluster problem. 

In addition to the differences mentioned 
above (detailed knowledge of the Hamiltonian is not needed for the HF 
algorithm so long as we have an initial Green function), there are other 
advantages to the HF algorithm.  Whereas in the BSS algorithm, all 
degrees of freedom must appear explicitly, in the HF algorithm, any 
non-interacting degrees of freedom may be integrated out without loss of 
any information.  At the end of the calculation, the irreducible diagrams 
on the interacting orbitals may be used to calculate any relevant quantity.  
Therefore, the HF algorithm may be used, for example, to simulate the 
periodic Anderson model (with only the f-orbital correlated) with the 
same computational cost required to simulate a single-band model.  One 
may also incorporate a (dynamical) mean field coupling to an 
environment, or between an infinite set of coupled Hubbard 
planes\cite{DCA_Maier_comment}, at no additional computational cost. 
In these cases, the information about the mean field coupling to the 
environment or the other planes is reflected in $\Gscript$.  

	For clusters, the Hirsch-Fye algorithm is very stable at low
temperatures.  In particular, the matrices $G_c^{\sigma}$ 
which are generated in the algorithm are well-conditioned, the costly 
stabilization steps required at low temperatures\cite{white2,Loh_92} for 
the more popular BSS algorithm are avoided. 
  
	Finally, the Hirsch-Fye algorithm is easily adapted to 
making measurements which are non-local in time, such as those
required to calculate the irreducible vertex functions.  This will
be discussed in the next section.  It is very difficult to measure
quantities which are non-local in time with the BSS algorithm.
In fact, such measurements require significantly more CPU
time than is required to average over the HHS field configurations,
since the CPU time required by these measurements scales like
$(N_c N_l)^3$ for both the BSS and Hirsch-Fye algorithms.  Thus,
when these measurements which are non-local in time
are required, both algorithms scale like $(N_c N_l)^3$.

\subsection {Making and Conditioning Measurements}
\label{SSec_QMCMEAS}

In the QMC technique, all the physical quantities are expressed in
terms of Green functions. Standard diagrammatic techniques are applied
to evaluate these quantities. In doing so one must remember that
the Hubbard-Stratonovich transformation reduces the problem to one
of free electrons moving in a time-dependent field. Thus for each
field configuration, any diagram may be formed by summing all allowed
Wick's contractions. The full quantity is recovered by averaging this 
over all field configurations. Connected as well as disconnected
configurations must be used during the evaluation. It is important to
average over all equivalent time and space differences and all the 
symmetries of the Hamiltonian in order to produce the lowest variance 
measurement.

One  difficulty encountered with the DCA algorithm is that a reliable
transform from imaginary-time quantities, in the QMC part, to Matsubara 
frequencies, for the coarse-graining part is needed.   A careful treatment 
of the frequency summation or the imaginary-time integration is crucial in 
order to ensure the accuracy and the stability of the algorithm and
to maintain the correct high-frequency behavior of the Green functions. 
We need to evaluate the following integral
\begin{eqnarray}
G_c({\bf K},i\omega_n) = 
\int_{0}^{\beta} d\tau e^{i\omega_n \tau}G_c({\bf K},\tau)  \quad\mbox{.} 
\label{FTG}
\end{eqnarray}
But from the QMC, we know the function $G_c({\bf K}, \tau)$ only at
a discrete subset of the interval $[0,\beta]$. As it may be readily seen
by discretizing the above equation, the estimation of $G_c({\bf K}, i\omega_n)$
becomes inaccurate at high-frequencies. This is formalized by Nyquist's theorem
which tells us that above the frequency $\omega_c= \frac{\pi}{\Delta \tau}$ 
unpredictable results are produced by conventional quadrature techniques.
For example, a rectangular approximation to the integral in Eq.~\ref{FTG} 
yields a $G(\K,i\omega_n)$ that is periodic in $\omega_n$. This presents a 
difficulty since the causality requires that
\begin{eqnarray}
\lim_{\omega_n \rightarrow \infty} G(\K,i\omega_n) \approx \frac {1}{i\omega_n}\,.
\label{limit}
\end{eqnarray}
 
A straightforward way to cure this problem may be to increase the size of the 
set of $\tau$-points where the Green function is evaluated. But, this renders 
the QMC simulation rapidly intractable as seen in the previous section. A 
more economic way to avoid the problem is to use the high frequency
information provided by an approximate method that is asymptotically exact.

Second-order perturbation theory is enough to obtain the correct asymptotic 
behavior, Eq.~\ref{limit}. To use this high frequency information, we compute 
the Matsubara-frequency Green function from the imaginary-time QMC Green 
function as 
follows\cite{jarrell}
\begin{eqnarray}
G_c({\bf K},i\omega_n) &=&  G_{c\,pt}({\bf K},i\omega_n) + \nonumber \\
                       & & \int_{0}^{\beta} d\tau e^{i\omega \tau}
(G_c({\bf K},\tau)-G_{c\,pt}({\bf K},\tau)) \quad\mbox{.} 
\end{eqnarray}
The integral is computed by first splining the difference
$G_c({\bf K},\tau)-G_{c\,pt}({\bf K},\tau)$ using an Akima spline\cite{akima}, 
and then integrating the spline (a technique often called oversampling). 

As another example, consider the local magnetic susceptibility (used to
calculate the screened local moment)
\begin{eqnarray}
\chi(T) & \approx &
\frac{1}{N_c} \sum_\i \int_0^\beta d\tau 
\left\langle S^+_\i(\tau) S^-_\i(0)\right\rangle \nonumber \\
&\approx &\frac{1}{N_c} \sum_\i \int_0^\beta d\tau 
\left\langle c_{\up \i}^{\dag}(\tau)c_{\dwn \i}(\tau)
c_{\dwn \i}^{\dag}(0)c_{\up \i}(0)\right\rangle \nonumber \\
&\approx & \frac{T}{2N_c} \sum_{\si \i} \int_0^\beta d\tau \int_0^\beta d\tau'
{\large{\langle}} \Gc_\si(\i,\tau+\tau';\i,\tau') \nonumber \\
& &\times \Gc_{-\si}(\i,\tau';\i,\tau+\tau') {\large{\rangle}}_{\{ s_{\i l} \}}
\label{localsus}
\end{eqnarray}
where the ${\{ s_{\i l} \}}$ subscript indicates that the Monte Carlo 
average over the Hirsch-Hubbard-Stratonovich fields is still to be 
performed, and in the last step in Eq.~\ref{localsus} we form all 
allowed Wick's contractions and average over all equivalent time and 
spatial differences to reduce the variance of this estimator.  This 
measurement is best accomplished by splitting it in two parts.  First, 
we measure 
$\chi(\tau)$
\begin{eqnarray}
\chi(\tau) &=& \frac{T}{2N_c} \sum_{\si \i} \int_0^\beta d\tau'
{\large{\langle}} \Gc_\si(\i,\tau+\tau';\i,\tau') \nonumber \\
& & \times \Gc_{-\si}(\i,\tau';\i,\tau+\tau')
{\large{\rangle}}_{\{ s_{\i l} \}}
\end{eqnarray}
by approximating the integral as a sum using a rectangular approximation.
For $\tau>0$
\begin{eqnarray}
\chi(\tau_l) &\approx & \frac{1}{2N_l N_c} \sum_{\si \i l'}
{\large{\langle}} \Gc_\si(\i,ind(l+l');\i,l') \nonumber \\
& & \times \Gc_{-\si}(\i,l';\i,ind(l+l'))
{\large{\rangle}}_{\{ s_{\i l} \}} \,,
\end{eqnarray}
where $ind(l)$ is the smaller nonnegative value of either $l$ or $l-N_l$.
For $\tau=0$ the fact that we always store
$G_{\si}(\i,l';\i,l')=G_{\si}(\i,\tau_{l'}+0^+;\i,\tau_{l'})$ requires 
us to modify the measurement
\begin{eqnarray}
\chi(\tau=0)\approx \frac{1}{2N_l N_c} && \sum_{\si,l',\i}
{\Large{\langle}} \Gc_\si(\i,l';\i,l') \nonumber \\
&&\left( \Gc_{-\si}(\i,l';\i,l')-1\right)
{\Large{\rangle}}_{\{ s_{\i l} \}}\,.
\end{eqnarray}
Finally
\begin{eqnarray}
\chi(T) = \int_0^\beta d\tau \chi(\tau)
\approx \sum_l f(l) \Delta\tau \chi(\tau_l)\,,
\end{eqnarray}
where the Simpson factor $f(l) = 2\Delta\tau/3$ ($4\Delta\tau/3$) for odd
(even) $l$ is used to reduce the systematic error of the integral.

        As a final example, consider the cluster particle-particle 
Green function matrix $\chi_c(q,K,K')$ ($K=(\K,i\omega_n)$)  
which is used in Sec.~\ref{SSSec_particle_particle} to calculate 
the lattice pair-field susceptibilities.  The first step is to form
the corresponding quantity in the cluster space-time
\begin{equation}
\chic(X_1,X_2,X_3,X_4)=
\left\langle T_{\tau} c_{\uparrow}(X_1)c_{\downarrow}(X_2)
          c_{\downarrow}^{\dagger}(X_3)c_{\uparrow}^{\dagger}(X_4)
\right\rangle\,.
\end{equation}
Here $X_i$ is in the space-(imaginary)time notation $X_i=(\X_\i,\tau_i)$,
where the points $\X_\i$ are on the corresponding reciprocal cluster of 
$\K$ in real space, and $\left\langle T_{\tau} ...\right\rangle$ denotes 
the time ordered averaging process.  The two-particle Green functions 
are difficult to measure efficiently.  For a particular configuration of 
the auxiliary Hubbard-Stratonovich fields, the fermions are noninteracting, 
thus the expectation value indicated above may be evaluated in two steps.  
First, using Wick's theorem, its value is tabulated for each field 
configuration $\{s_i\}$ and then transformed into the cluster Fourier 
space.  Second, we Monte Carlo average over these configurations.  After 
the first step, the expression for the above two-particle Green function 
in the cluster momentum-frequency space becomes
\begin{eqnarray}
&&\chic(\Q,i\nu_n;\K,i\omega_n;\K',i\omega_{n'}) = \nonumber \\
%\left\langle
{\LARGE{\langle}}
&&\sum_{X_1,X_4} e^{iK'X_1}\Gc_{\uparrow}(X_1,X_4)e^{-iKX_4} \nonumber \\
&&\sum_{X_2,X_3} e^{i(Q-K')X_2}\Gc_{\downarrow}(X_2,X_3)e^{-i(Q-K)X_3}
%\right\rangle
{\LARGE {\rangle}}_{{\{ s_i \}}}\,.\label{chi2}
\end{eqnarray}
where $K$ is the momentum-frequency point $K=(\K,i\omega_n)$.  The 
average over Hubbard-Stratonovich fields 
$\left\langle ...\right\rangle_{\{ s_i \}}$ can be evaluated through 
the QMC sweeps along with the measurements of $\Gc_{\uparrow}$ and 
$\Gc_{\downarrow}$.  However, the sums (integrals) over $\tau$ 
in Eq.~\ref{chi2} require special consideration.  Since the Green functions
change discontinuously when the two time arguments intersect, the best 
applicable integral approximation is the trapezoidal approximation.  Using 
this, we will run into Green functions $\Gc(\X,\tau;\X,\tau)$ with both 
time and space arguments the same. In the QMC algorithm, this is stored 
as $\Gc(\X,\tau^+;\X,\tau)$  (i.e.\ it is assumed that the first time 
argument is slightly greater than the second); however, if we replaced 
the equal time Green function to be the average 
$\{\Gc(\X,\tau^+;\X,\tau)~+\Gc(\X,\tau;\X,\tau^+)\}/2~=
\Gc(\X,\tau^+;\X,\tau)-1/2$ 
then a trapezoidal approximation of the integrals results.  If we call the 
matrix ${\bf G}_c$, with $1/2$ subtracted from its diagonal elements, as 
${\hat \G}_c$ (note that we can treat one of the three independent momenta 
involved in $\chic$ as a variable $Q$ outside the matrix structure), then 
we can write the two-particle Green function in a matrix form  
\begin{eqnarray}
\label{chi_estimate}
&&\chic_{\imath \jmath}(\Q)= \\
&&\left\langle
\left( \F_{\Q=0}^{\dagger} {\hat \bGc}_{\uparrow} \F_{\Q=0} \right)_{\imath \jmath}
\left( \F_{\Q}^{\dagger} {\hat \bGc}_{\downarrow} \F_{\Q} \right)_{\imath\jmath}^*
\right\rangle_{{\{ s_{i} \}}}\nonumber \,,
\end{eqnarray}
where $(\F_{\Q})_{i\jmath}= 
\Delta\tau e^{-i(\K_\jmath-\Q){\bf .}\X_i-i\omega_\jmath\tau_i}$
where we have chosen $\imath$ and $\jmath$ to index the cluster 
momentum-frequency space.
This measurement may be performed efficiently if the product of three 
matrices in each set of parenthesis is tabulated as two sequential 
matrix-matrix products and stored before the direct product between 
the terms in parenthesis is calculated.  When done this way, the 
calculation time required for this process  scales like $(N_cN_l)^3 $ 
rather than $(N_cN_l)^4$ as would result from a straight-forward 
evaluation of the sums implicit in  Eq.~\ref{chi_estimate}.
    
%---------------------------------------------------------------

For the reasons discussed above, Eq.~\ref{chi_estimate} becomes 
unreliable at high frequencies $|\omega_n| > \pi/\Delta\tau$.  
The high frequency behavior of the two particle Green function 
can be recovered by using a method similar to that developed for 
the one particle Green function\cite{deisz}.  The first term of 
its perturbation expansion, the bubble diagram, is used for the 
conditioning. It is calculated in two ways: First it is formed 
from the square of the properly conditioned cluster Green function.  
Second, it is calculated using the same approximation to the Fourier 
transform employed in Eq.~\ref{chi_estimate}.  The difference of 
the two may be used to condition the estimate
\begin{eqnarray}
&&\chic_{\imath\jmath}(\Q)=\nonumber \\
&&\left\langle
\left( \F_{\Q=0}^{\dagger} {\hat \bGc}_{\uparrow} \F_{\Q=0} \right)_{\imath\jmath}
\left( \F_{\Q}^{\dagger} {\hat \bGc}_{\downarrow} \F_{\Q} \right)_{\imath\jmath}^*
\right\rangle_{{\{ s_i \}}}-\nonumber \\
&&\left( \F_{\Q=0}^{\dagger} \left\langle {\hat \bGc}_{\uparrow}
\right\rangle_{{\{ s_i \}}} \F_{\Q=0} \right)_{\imath\jmath}
\left( \F_{\Q}^{\dagger} \left\langle {\hat \bGc}_{\downarrow}
\right\rangle_{{\{ s_i \}}} \F_{\Q} \right)_{\imath\jmath}^* 
\nonumber \\
&& + \Gc( K_\imath ) \Gc^*( K_\imath-Q )\delta_{\imath\jmath}\,.  \nonumber \\
\label{chi_betterestimate}
\end{eqnarray}
Moreover, this appends the right asymptotic behavior of the perturbation 
result to the two-particle Green function at high frequencies where QMC
results are dominated by statistical errors.

\subsection{Optimizing the Code}
\label{SSec_DCA_code}

        In this subsection, we will discuss the optimization
and parallelization of the QMC code.

	We generally find that the heat bath algorithm is more efficient,
presumably because it has a lower acceptance rate and therefore
deemphasizes the expensive step of updating the Green function.

	We may greatly reduce the statistical error in many of
the measured Green functions by employing the translational and
point-group symmetries of the cluster.  The QMC averaging over
the HHS fields systematically restores the translational invariance 
of the system in time and space.  So we may reduce the statistical
error in the measured Green functions by averaging over all equivalent
differences in spatial and temporal cluster coordinates.
To reduce the statistical error further, we then average over all 
the lattice point group operations.  For example, for $G_c(\K,\omega_n)$
\begin{equation}
G_c(\K,i\omega_n) = \frac{1}{N_\R} \sum_\R G_c\left(\R(\K),i\omega_n\right)
\end{equation}
where $\R$ is one of the symmetry operations in the point group of the 
lattice and $N_{\R}$ is the total number of such symmetry operations.

The two-particle Green functions typically have more statistical noise than 
their single-particle counterparts, and their matrices can be quite large.  
To reduce both the storage needed for these measurements and their 
statistical noise, the point group symmetry of the lattice may again 
be used.  We first average the two-particle cluster Green functions 
over the different point-group operations
\begin{equation}
\chi_{\si,\si'}(q,\K,\K') = \frac{1}{N_{\R}}
\sum_{\R} \chi_{\si,\si'}\left(q,\R(\K),\R(\K')\right) \,.
\label{symsuscep} 
\end{equation}        
We should also average over the symmetries of the diagrams.  I.e., for 
the particle-particle channel there are additional symmetries of the 
diagrams which include horizontal $(\K,i\omega_n;\K',i\omega_{n'})
\rightarrow (\K',i\omega_{n'};\K,i\omega_n)$ and vertical
$(\K,i\omega_n;\K',i\omega_{n'}) \rightarrow  (-\K,-i\omega_n;-\K',-\omega_{n'})$
reflections.   After these symmetries have been imposed, we will lose 
no information and significantly reduce the storage requirements
if we store $\chi_{\si,\si'}(q,\K,\K')$ for either $\K$ or $\K'$
within the irreducible wedge (we may not take both $\K$ and $\K'$ 
within the irreducible wedge though).

	The memory required for these calculations may be further
minimized by limiting the use of the double precision arithmetic.  
The Green functions and all of the equations associated with the 
calculation of the initial Green function and the Green function 
update, Eq.~\ref{updateG}, are computed with double precision (8 
byte real) to minimize the problems with the accumulation of 
numerical error discussed in Sec.~\ref{SSec_QMC_ALG}.  However, 
to save memory, it is convenient to both calculate and store the 
two-particle cluster Green functions with single precision (8 byte 
complex).  Since these measurements typically have a fraction of 
a percent statistical error, higher precision arithmetic and 
storage will not improve the accuracy of the two-particle measurements.

The required CPU time may be reduced by optimizing the inner loops.
The two numerically most expensive parts of the QMC code are
the Green function update, Eq.~\ref{updateG}, and the two-particle
measurements, Eq.~\ref{chi_betterestimate}.  These can be written 
in terms of highly-optimized BLAS calls\cite{netlib}, DGER and CGEMM, 
respectively.  To see that Eq.~\ref{updateG} can be calculated with 
an outer product, we define
\begin{equation}
a_{m} = \frac{(e^{-\alpha\sigma (s_m-{s'}_m)}-1)}{1+(1-\Gc_{\sigma m,m})
                (e^{-\alpha\sigma (s_m-{s'}_m)}-1)}\,.
\end{equation} 
Then Eq.~\ref{updateG} takes the form
\begin{equation}
{{\Gc}'}_{\sigma i j} = \Gc_{\sigma i j} +
(\Gc_{\sigma i m}-\de_{i,m})a_{m}  \Gc_{\sigma m j}\,.
\end{equation} 
This is a vector outer product and matrix update, with vectors 
$a_{m} (\Gc_{\sigma i m}-\de_{i,m})$ and $\Gc_{\sigma m j}$
for fixed $m$.

Additional speedup of the calculation is possible by writing parallel 
codes.  The DCA--QMC codes are extremely well suited for massively 
parallel supercomputers because of their efficient use of the 
floating-point capabilities of such machines and the highly parallel 
nature of the codes and the underlying algorithm.  With the current 
relative decline in the availability of vector supercomputers
and the concomitant increase in the number of massively parallel
supercomputers, this is an important feature of the algorithm.
In the remainder of this subsection, we discuss first the general 
parallel nature of the algorithm. 
% In fact, we have written
%two different parallel versions of our DCA code in order to most
%efficiently use the computational resources of various parallel
%architectures for various DCA cluster sizes and temperatures.

There is a high degree of parallelism in the DCA--QMC algorithm, 
which one may exploit.  This parallelism exists on two levels.  
First, QMC is itself inherently parallel because it consists of a 
number of stochastic random-walks.  One may think of QMC as one 
long Markov-chain walk.  Measurements are made periodically along 
this walk.  At the end of the walk, these measurements are averaged 
and the final result, with error bars, is obtained.

However, there is no reason why this Markov-chain walk has to be
continuous.  It has been known for years that one can perform several 
independent, shorter Markov-chain walks and average the results of 
each walk to obtain the final result of the calculation.  The result 
can be an almost perfect parallel speedup as an increasing number of 
processors is applied to a problem.  This arises because only an 
extremely small amount of communication between processors is 
required -- first to initialize the Markov-chain walks and then to 
collect the data for averaging at the end of the Markov process 
(even this averaging can be done in parallel using MPI calls).  We 
call this the ``Perfectly Parallel'' algorithm.

The second degree of parallelism exists in the linear algebra problem
itself.  That is, one can distribute the vectors and matrices which
comprise the linear algebra problem across several processors.  (The
matrix in our case is the Green function discussed above.)  Such
a break-up of the data becomes of paramount importance when the size
of a matrix is so large that it cannot possibly fit within all of the
memory available on a single processor of a computer.  

The issue of interprocessor communication now becomes paramount as
one performs linear algebra.  However, two things work in our favor
here.  First, the main linear algebra operation of the QMC
is a vector outer product -- which is in itself inherently parallel.
Second, this is a well-studied problem and again an efficient 
library package, the parallel PBLAS\cite{netlib}, exists to solve it.  
When we divide the Green function over all of the processors
that we use in a run on a parallel machine, 
we call this the ``Truly Parallel'' method.

The Truly Parallel method can be used to efficiently fill all
available processors of a parallel machine with one DCA--QMC problem.  
Often, however, the problem of interest is not so big as to require 
the entire machine for one Green function, but is too big to fit within 
the RAM available on a single processor and hence too big for the 
Perfectly Parallel code.  To efficiently use available hardware for 
these problems, one can employ a ``Hybrid'' code, which is both truly 
parallel in part and perfectly parallel in part.

The hybrid code may be thought of as using blocks of processors to
distribute Green functions and in turn performing a perfectly parallel
QMC with many such blocks.  For example, say that the Green function
for the problem at hand will not fit in the memory of a single
processor, but will fit within the memory of 4 processors.  Assume also 
that there are 100 processors available for a run.  The hybrid code 
then allocates all 100 processors, divides these 100 processors into
25 blocks of 4, distributes copies of the initial Green function onto 
each of the 25 blocks, and then does a perfectly parallel QMC using 
these 25 blocks.  This makes the most use of the resources of a machine 
and is especially well-suited for a Symmetric Multi-Processor machine, 
where many nodes exist and each node comprises several processors with 
a shared, relatively large, pool of RAM.

\section{The DCA algorithm}
\label{DCA_algo}

\begin{figure}[ht]
\leavevmode\centering\psfig{file=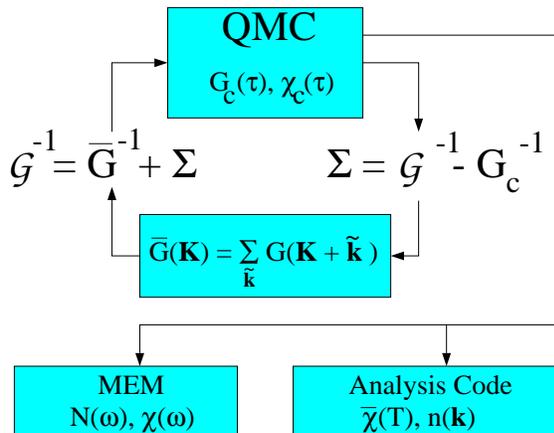,width=3. in}
\caption{Sketch of the DCA algorithm: the self-consistent loop,
the analysis part and the MEM part.}
\label{algorithm}
\end{figure}

        In this section, we will discuss how the QMC and DCA formalism
are combined into a DCA algorithm for simulating lattice problems. The 
complete DCA program is made of three completely separate parts as 
illustrated in Fig.~\ref{algorithm}. The first part is the self-consistent 
loop which is the main part of the algorithm. It includes  the DCA 
self-consistency loop composed of the QMC block and the coarse-graining 
of the lattice.  This program is usually run on a parallel supercomputer,
and the cluster self energy and the various two-particle cluster
Green functions are written into files. In the second part various
 one-particle and two-particle lattice Green functions are 
calculated from the cluster Green functions obtained from the 
self-consistent loop. This part of the code is generally run on a 
workstation and it requires in the data generated by the first part of
the code.  The third  is devoted to the analytical continuation of the 
imaginary-time Green functions to real frequencies using 
the Maximum Entropy Method (MEM).
 
\subsection{Part 1: The self-consistent loop}
 
\begin{enumerate}

\item{The DCA iteration procedure  is started by setting the initial 
self-energy $\Sigma_c(\K,i\omega_n)=0$, or to a perturbation theory
result.}

\item{ $\Sigma$ is then used to compute  the coarse-grained Green 
function ${\bar {G}}(\K,i\omega_n)$,
\begin{eqnarray}
\bar {G}(\K,i\omega_n)=\sum_{\tk} \frac{1}{i\omega_n-\ep-\epsilon_{\K+ \bf \tk}-
  \Sigmac(\K,i\omega_n)}\,.
\label{coarse2}
\end{eqnarray} 
}

\item{ The next step of the iteration is to use ${\bar G}(\K,i\omega_n)$ 
to compute the host Green function 
${\cal{G}}(\K,i\omega_n)^{-1}=\bar{G}(\K,i\omega_n)^{-1}+
\Sigmac(\K,i\omega_n) $ which must be introduced to avoid over-counting
diagrams.  ${\cal {G}}(\K,i\omega_n)$ serves as the input to the QMC 
simulation to yield a new estimate for the cluster self-energy.}

\item{${\cal {G}}(\K,i\omega_n)$ must be Fourier transformed from 
the momentum-frequency variables to space-imaginary-time variables 
before being introduced in the QMC part of the program as the 
initial Green function $\Gc_{ij}=\Gscript(\X_i-\X_j,\tau_i-\tau_j)$ 
corresponding to all $s_i=0$.  Eq.~\ref{updateG} is used to 
generate the cluster Green function corresponding to $s_i=1$,
or to the $\{s_i\}$ from a previous run.
}  

\item{The QMC step is next and is the most time consuming part of 
the algorithm. Each QMC step is warmed up before one starts to 
perform measurements.  While making measurement, we average over
the differences in space and time and the point group operations, 
as described above, to reduce the statistical error. This together
with the QMC averaging restores the translational invariance of the 
system in time and space, so 
$\left\langle\Gc_{ij}\right\rangle_{\{s_i\}}=\Gc(\X_i-\X_j,\tau_i-\tau_j)$.}

\item{ $\Gc(\X_i-\X_j,\tau_i-\tau_j)$ is then Fourier-transformed to 
$G_c(\K,i\omega_n)$.  We calculate a new estimate for the self-energy 
$\Sigmac(\K,i\omega_n)={\cal{G}}(\K,i\omega_n)^{-1}-\Gc(\K,i\omega_n)^{-1}$.
}

\item{Starting with step 2, the procedure is repeated until 
$\Sigmac(\K,i\omega_n)$ converges.  This typically happens in less than ten 
iterations. The number of iterations decreases when $N_c$ increases since 
the coupling to the host is smaller (${\cal O}(1/N_c)$) \cite{DCA_Maier1} 
for larger clusters.
  
The convergence test is made on the ratio $\rho$,
\begin{equation}
\rho=\frac{|\sum_\K (\Sigmac_{new}(\K,i\omega_0)-\Sigmac_{old}(\K,i\omega_0))|}
{|\sum_\K \Sigmac_{old}(\K,i\omega_0)|} 
\end{equation}
where $\omega_0=\pi T$
} 

\item{Once convergence is reached to the desired accuracy, the remaining 
one and two-particle measurements are made in a final QMC iteration. 
As in a usual QMC simulation, bins of measurements are accumulated 
and error estimates are made from the fluctuations of the binned
measurements.  These error estimates are accurate provided that 
the bins contain enough measurements so that the bin-averages are 
uncorrelated.  The statistical error may be reduced by averaging
over the different symmetries as discussed above. 
}

\end{enumerate}

Once the cluster Green functions are obtained, the determination of 
the lattice quantities requires additional steps which are done in 
separate programs.  

\subsection{Part 2: Numerical calculation of lattice quantities}

The self-consistent loop yields cluster Green functions 
$G_c(\K,i\omega_n)$, $\Sigmac(\K,i\omega_n)$,
and susceptibilities 
$\chic_{\sigma,\sigma'}(\K,i\omega_n;\K',i\omega_m)$, 
$\Pi_g(\K,i\omega_n;\K',i\omega_m)$ 
which may be used to construct the equivalent lattice quantities.
This is done in a separate computer program in which the irreducible
quantities of the cluster which are in the DCA approximation
identical to those of the lattice are used to compute the
corresponding reducible lattice quantities. 

To calculate the single-particle quantities, an interpolated self-energy 
$\Sigma(\k,i\omega_n)$ may be used.  This is especially important for the 
calculation of $\left |\grad n(\k)\right|$, and other quantities 
such as band structure, where continuity of the self energy is important.  
We often use bilinear interpolation for this purpose, since it is 
guaranteed to preserve the sign of function (i.e. the bilinear 
interpolation of a positive-definite function remains everywhere 
positive).  We also use a multidimensional spline interpolant, like some 
Akima splines, which does not overshoot.  However, it is important to note 
that this interpolated self-energy should not be used in the self-consistent 
loop as this can lead to violation of causality\cite{DCA_Hettler2}. 

The interpolated self energy, $\Sigma(\k,i\omega_n)$ is then used to 
calculate the Fermi surface. For this we use the discrete form of 
$\grad n(\k)$   
\begin{eqnarray}
\frac{\Delta n_{\k}}{\Delta \k}=\frac{T \sum_{n} G(\k+\Delta \k,i\omega_n)-G(\k,i\omega_n)}
 {\Delta \k}
\label{dmomentum2}
\end{eqnarray}
which, in a Fermi liquid (or a marginal Fermi liquid) is maximum at the
Fermi surface.  The quasiparticle weight may be approximated with
\begin{eqnarray}
Z_{\k}\approx 1-\frac{Im \Sigma(\k,i\omega_{n=0})}{\omega_{n=0}}
\label{quasipart2}
\end{eqnarray}                   
which becomes exact as $T \rightarrow 0$. 

The calculation of the lattice susceptibilities in the particle-hole
channel and in the particle-particle channel is also made in this code.  
The stored QMC cluster susceptibilities are used for this purpose as 
prescribed in Sec.~\ref{SSec_twopart}.  Here, we first form the 
corresponding coarse-grained and bare cluster susceptibilities, and
then we use Eq.~\ref{chicoars2} to calculate the corresponding
coarse-grained lattice susceptibility.  To calculate the pair field
susceptibilities, we first calculate the corresponding coarse-grained
two-particle reducible vertex, and then use Eq.~\ref{pair_dyson} to 
calculate the coarse-grained lattice pair-field susceptibility matrix.  

In the two-particle calculations, it is tempting to interpolate the 
cluster vertex functions to the lattice momenta.  However, this 
would increase the size of the matrices which must be inverted in 
Eq.~\ref{chicoars2} dramatically, making the calculation of the
lattice susceptibilities numerically much more expensive.

\subsection{Part 3: Analytic continuation}

Unfortunately, there is no reliable way to perform the direct analytic
continuation of $\Sigmac({\K},i\omega_n)$.  Pad\'e approximants lead 
to very unstable spectra because of the QMC statistical noise contained 
in $\Sigmac(\K,i\omega_n)$.  The binned imaginary-time Green function 
data accumulated from the cluster calculation must be used to obtain 
lattice spectra from which $\Sigmac({\bf K}, \omega)$ may be deduced.  
To obtain the self-energy and spectral-weight function $A(\bf k, \omega)$ 
of the lattice in real frequencies, we first compute the cluster 
spectral-weight $\bar{A}(\bf K, \omega)$.   This is done using the Maximum 
entropy method \cite{JARRELLandGUB} to invert the following integral 
equation
\begin{eqnarray}
\bar {G}({\bf K}, \tau) = \int d\omega \frac  {e^{-\omega \tau}}{1+
e^{-\beta \omega}} \bar {A}(\bf K, \omega)  \quad\mbox{,}
\label{mem}
\end{eqnarray}
where $\Gbar({\bf K}, \tau)$ is the imaginary-time Green function obtained
from the QMC simulation of the cluster. 

Since $\bar A(\K,\omega)=-1/\pi Im\Gbar(\K, \omega)$, the full 
frequency-dependent coarse-grained Green function $\Gbar(\K,\omega)$ is 
obtained using Kramers-Kronig relations. Then,  the equation
\begin{equation}
\Gbar(\K,\omega)= \frac{N_c}{N}\sum_\kt \frac{1}{\omega-\ep-\ep_{\K+\kt}-\Sigmac(\K,\omega)}
\end{equation}
is solved for the real-frequency self-energy $\Sigmac(\K,\omega)$ using a 
complex root finder\cite{numrec_root}. This self energy may then 
be inerpolated onto the lattice $\k$ points using a high-level
interpolant which also preserves the sign of the imaginary 
part.

The above steps are unnecessary if the local quantities are to
be computed since the local lattice and cluster Green functions
correspond one-to-one.  For example, we may directly analytically continue 
the local cluster Green function to obtain the lattice density
of states.

\section{Application to the 2D Hubbard model}
\label{Sec_Results}

        We will now show the results of the application of the DCA to 
the two-dimensional Hubbard model. The Hubbard model has a long history 
and is believed to contain the mechanism of various physical phenomena 
such as magnetism, metal-insulator transitions and more recently 
superconductivity and non-Fermi liquid behavior.  Our intent in this 
section is not to exhaustively study this model's properties, but 
rather to use it to illustrate the power and limitations of the DCA 
and to survey what can be done.

        Since the two-dimensional model is not expected to have a
finite-temperature magnetic or perhaps even superconducting transition, we 
will add a hopping $t_\perp$\cite{tperp} into the third dimension between 
an infinite set of weakly coupled Hubbard planes    
\begin{equation}
t_\perp (k_x,k_y,k_z)=-2t_\perp (\cos k_x-\cos k_y)^2\cos k_z \,.
\end{equation}
We take $t_\perp\ll t$, and treat the additional coupling at the DMFA 
level, so the self-energy is independent of $k_z$.  This is accomplished 
by modifying the coarse-graining cells into rectangular solids of 
dimensions $\Dk$, $\Dk$ and $2\pi$ in the $k_x$, $k_y$ and $k_z$ 
directions, respectively.  After coarse-graining, the problem is
reduced to a two-dimensional cluster.  Information relevant to the 
mean-field coupling between the planes is contained within $\Gscript$.

\subsection{Results at Half-filling}

        The physics of the half filled model is a severe test of
the DCA as well as finite-sized simulations (FSS) due to the quantum 
critical point at zero doping.  As this point is approached, both the 
dynamical and spatial correlation lengths diverge, and both the DCA 
and FSS are expected to fail.
                                                                       
\subsubsection{Antiferromagnetism}

Earlier finite size simulations \cite{hirsch,white} employing the QMC 
method have led to the conclusion that the ground state is an 
antiferromagnetic insulator at half-filling.  Since the model is two 
dimensional, we know from the Mermin-Wagner theorem that the transition 
temperature
is necessarily zero. But as found in infinite dimensions\cite{jarrell}, 
the DMFA predicts a finite temperature transition even in two dimensions. 
This spurious behavior may be attributed to the lack of non-local 
correlations in the DMFA. These correlations are known to induce strong 
fluctuations particularly in reduced dimensions and are responsible for 
the suppression of the finite temperature transition. The DCA which 
includes these non-local correlations is thus expected to progressively
drive the spurious finite temperature transition found in the DMFA towards 
zero temperature as the cluster size increases.

\begin{figure}[ht]
\leavevmode\centering\psfig{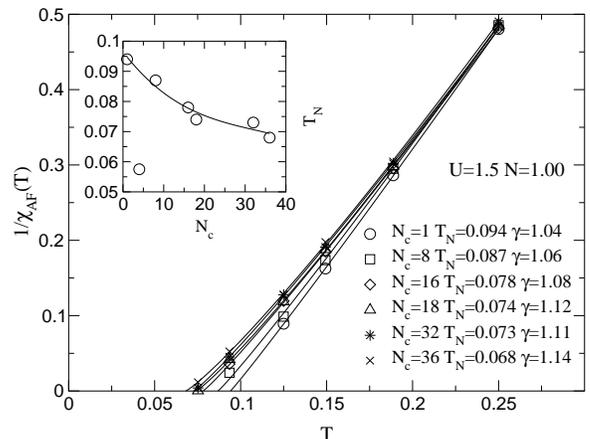}
\caption{The inverse antiferromagnetic susceptibility versus
temperature for various cluster sizes.  The lines are fits to
the function $(T-T_{\rm{N}})^\gamma$.  In the inset, the corresponding 
Ne\'{e}l temperatures, determined by the divergence of the 
susceptibility, are plotted.  The line is a polynomial fit to 
the data, excluding $N_c=4$.}
\label{FIT_AF}
\end{figure}   

This behavior is illustrated in Fig.~\ref{FIT_AF}, where the inverse 
antiferromagnetic susceptibility is plotted versus temperature for 
$\delta=0$ and various values of $N_c$ which preserve the lattice 
symmetries as discussed in Sec.~\ref{SSec_Cells}. At high temperatures,
the susceptibility is independent of $N_c$, due to the lack of
non-local correlations.  In contrast to FSS calculations, the low
temperature susceptibility diverges at $T=T_{\rm{N}}$, indicating an 
instability to an antiferromagnetic phase. As $N_c$ increases $N_c>1$, 
the non-local dynamical fluctuations included in the DCA suppress the 
antiferromagnetism.  For example, when $N_c=1$, the susceptibility 
diverges with an exponent $\gamma\approx 1$, as expected for a 
mean-field theory; whereas the susceptibilities for larger $N_c$ values  
diverge at lower temperatures with larger exponents indicative of 
fluctuation effects\cite{caveat_gamma}.  At first, these effects are 
pronounced; however, as $N_c$ increases, $T_{\rm{N}}$ falls and $\gamma$ 
rises more slowly 
with increasing $N_c$.  This can be understood from the singular nature 
of the spin correlation length, which at least in the large $U$ limit is 
expected to vary as $\xi \propto e^{A/T}$, where $A$ is a constant of 
the order of the exchange coupling.  For this quantum critical transition, 
we expect the DCA to indicate a finite temperature transition once $\xi$ 
exceeds the linear cluster size.  Since correlations build exponentially, 
large increases in the cluster size will only reduce the DCA transition 
temperature logarithmically.

Note that the data for $T_{\rm{N}}(N_c)$ falls on a smooth curve, except for
$T_{\rm{N}}(N_c=4)$.  This behavior was seen previously in the transition
temperature of the Falicov-Kimball model, calculated with 
DCA.\cite{DCA_Hettler1,DCA_Hettler2}
The $N_c=4$ data falls well off the curve produced by the other data, 
and has a much larger exponent indicating that fluctuation effects are 
more pronounced.  Presently this behavior is not completely understood but 
may be related to the fact that the maximum coordination number for $N_c=4$ 
is two, whereas it is greater than two for cluster sizes larger than $N_c=4$. 

\begin{figure}[ht]
\leavevmode\centering\psfig{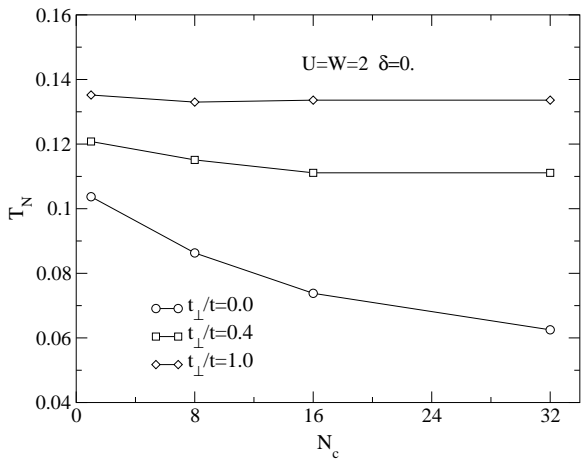}
\caption{$T_{\rm{N}}$ versus $N_c$ for different values of $t_\perp/t$}
\label{TNvsNc_tperp}
\end{figure}   

An interplanar coupling can significantly alter the phase diagram.
However, since the superexchange coupling varies roughly like the 
square of the hopping, it is necessary to make $t_\perp$ a significant
fraction of the intraplanar coupling $t$ in order to see an effect.
For example, if $t_\perp/t =0.4$, the ratio of the interplanar
to intraplanar exchanges is roughly $J_\perp/J \approx 0.16$.  
In Fig.~\ref{TNvsNc_tperp} the antiferromagnetic transition temperature 
is plotted versus $N_c$ when $t_\perp/t=0,0.4,1.0$ when $U=W=2$
and $\delta=0$.  For both $t_\perp/t=0.4$ and $t_\perp/t=1.0$,
the transition temperatures for $N_c=16$ and $32$ are the same
to within the numerical error.  Thus, the finite-temperature transitions 
found for small clusters, can be preserved as $N_c\to \infty$ by 
introducing the interplanar coupling.

\subsubsection{Mott transition at half-filling}

In the strong coupling limit, a Mott Hubbard gap is expected to 
open in the charge excitation spectra.  In the weak coupling limit,
the situation is less clear.  Since the ground state of the half filled
model is always an antiferromagnet, the system remains insulating, but 
the nature of the insulating state in weak coupling is less clear, and 
depends upon the dimensionality.  In one dimension, Lieb and Wu\cite{lieb} 
showed long ago that a charge gap opens as soon as $U > 0$. There is a 
spin-charge separation and there is no long range order, even at $T=0$. 
Hence, the Slater scenario is not responsible for the metal-insulator
transition and the low energy spin excitations are described by 
the Heisenberg model.  In infinite dimension, the model can be mapped to 
a self-consistent Anderson impurity problem. The solution of the 
self-consistent equations have been obtained numerically by various 
authors.  For small $U\ll W$, the antiferromagnetic transition temperature 
$T_{\rm{N}}$ is higher than any temperature at which there is a 
metal-insulator transition in the paramagnetic phase. Hence, the 
metal-insulator transition in infinite dimension is due to the Slater 
mechanism.  In two dimensions, the Mermin-Wagner theorem prohibits long 
range order for any $T > 0$, and we find that the weak coupling transition 
is similar to what is found in one dimension.

\begin{figure}[ht]
\leavevmode\centering\psfig{file=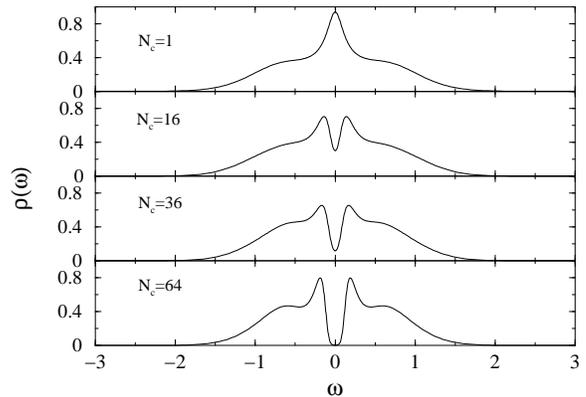,width=3. in}
\caption{The single-particle density of states $\rho(\omega)$ at 
$\beta=32$, $U=1$, and $t_\perp=0$.}
\label{doscu1}
\end{figure}
The density of states $\rho(\omega)$ (shown in Fig.~\ref{doscu1}) 
confirms the destruction of the Fermi liquid quasi-particle peak by 
short-range antiferromagnetic correlations.  With increasing $N_c$, the 
gap opens fully, and the Hubbard side bands become more pronounced. 
                     
\begin{figure}[ht]
\leavevmode\centering\psfig{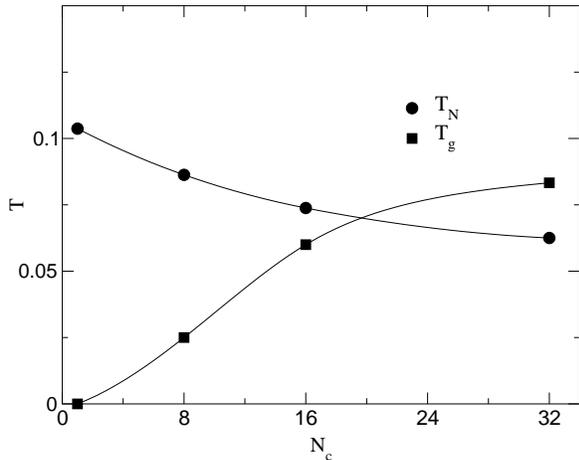}
\caption{The Neel, $T_{\rm{N}}$, and gap, $T_g$, temperatures versus 
$N_c$ when $U=W=2$, $\delta=0$ and $t_\perp=0$.}
\label{tng}
\end{figure}

In Fig.~\ref{tng} the behavior of $T_{\rm{N}}$ is compared to the 
temperature $T_g$ where the gap opens.  In contrast to $T_{\rm{N}}$, 
$T_g$ increases with the size of the cluster.  This confirms the 
conclusion that a gap, which is not due to antiferromagnetism alone,
opens at finite temperatures in the 2D Hubbard model. 

\subsubsection{ Comparisons with finite system simulations}

A great number of simulations have been performed in the half-filling 
regime of the two-dimensional Hubbard model over the last decade.  Most 
of them are based on QMC with imaginary-time data analytically continued 
by the maximum entropy method. While these studies all agree for $U > W$, 
for $U < W$ they have led to conflicting results\cite{DCA_Moukouri2}.  
The reason is that the metal-insulator transition is related to the 
antiferromagnetic transition so that it is difficult to distinguish 
between the two physical processes.  As a consequence various 
conflicting scenarios for the disappearance of the quasiparticle 
peak at low temperatures have been proposed. These controversies are 
inherent to the limitations of the finite system simulations. There 
are artificially introduced finite size gaps when the correlations 
become comparable to the system size. This does not occur when $U > W$ 
because the Mott gap opens well before the magnetic correlations set 
in. It is thus fair to ask to what extent the conclusion reached with 
the DCA above may be more reliable. For this it is necessary to compare 
the DCA to FSS.

\begin{figure}[ht]
\leavevmode\centering\psfig{file=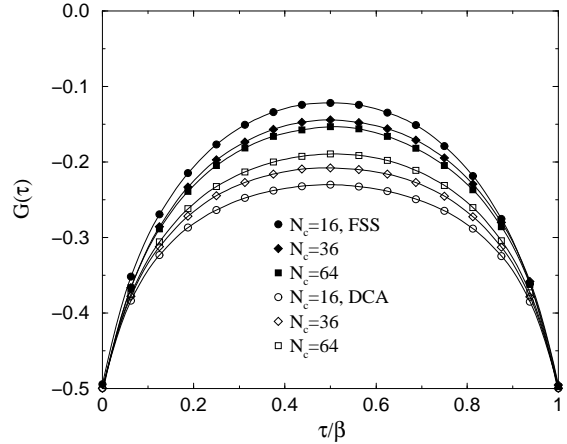,width=2.9 in}
\caption{The imaginary-time Green function at the point $X=(\pi,0)$ on 
the Fermi surface from finite size QMC (filled symbols) and from DCA 
(open symbols) with $U=1.1$, $\beta=16$, $t_\perp=0$ and $Nc=16, 36, 64$.
The size increases from top to bottom for FSS.  The size increases
from bottom to top for DCA.}
\label{gtau}
\end{figure}

\begin{figure}[ht]
\leavevmode\centering\psfig{file=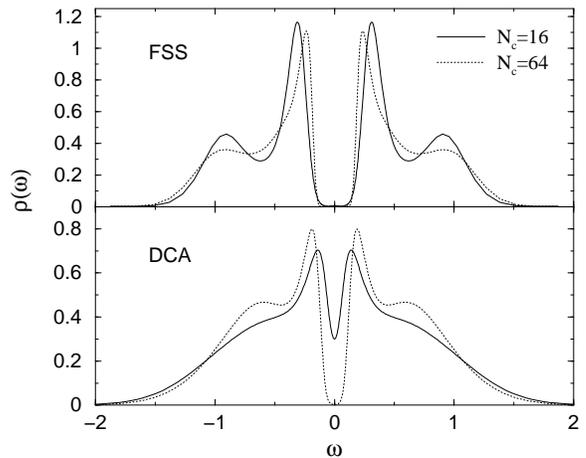,width=3. in}
\caption{  The density of states $\rho(\omega)$ from finite size QMC (top)
and from DCA (bottom) at $U=1$, $\beta=32$, $t_\perp=0$ and $Nc=16, 64$.}
\label{doscf}
\end{figure}

In Fig.~\ref{gtau} and Fig.~\ref{doscf}, we show the imaginary-time 
Green function $G(\tau)$ at the Fermi point $X=(\pi,0)$. This quantity 
has a more rapid decay from its maximum at $G(\beta/2)$ when the effects 
of the correlations are stronger. In finite systems, the decay is sharper 
for smaller lattices while in the DCA it is the opposite. This behavior 
marks the fundamental difference between the FSS and the DCA. At low 
temperatures, in FSS, the correlation length is greater than the lattice 
size. Thus, the effects of the correlations are overestimated for smaller 
clusters because these systems are artificially closer to criticality than
a system in the thermodynamic limit. This tendency is reduced by 
increasing the cluster size, which moves the system in the direction
of the thermodynamic limit. The situation is radically different
in the DCA where the system is already in the thermodynamic limit.
The DCA approximation consist in restricting correlations to within the
cluster length in the infinite system. As the cluster size increases,
possible longer-range correlations are progressively included. Thus, the 
effects of the correlations increase with the cluster size.

The density of states shown in Fig.~\ref{doscf} supports these conclusions. 
The finite size gap in the FSS decreases when the cluster size goes from 
$N_c=16$ to $64$. While, in the DCA for $Nc=16$ there is a pseudogap 
that turns into a true gap when the cluster size is increased to $64$.
Since by construction, the DCA underestimates the gap, we can affirm
that at this temperature, the gap exists in the thermodynamic limit. 
Its actual value is bracketed by the FSS and the DCA.  This behavior 
is characteristic of the DCA. It has been extensively verified on the 
one-dimensional Hubbard model\cite{DCA_Moukouri1}.

\subsection{Results away from half-filling}

\subsubsection{Sign Problem}

\begin{figure}[ht]
\leavevmode\centering\psfig{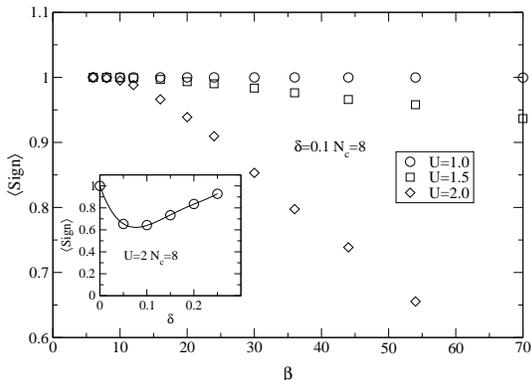}
\caption{The average sign as function of the inverse temperature $\beta$ 
for $N_c=8$ at $\delta=0.1$ for $U=1.0,1.5,2.0$.  In the inset, the average 
sign is plotted versus doping $\delta$ when $U=W=2$, $t_\perp=0$ and 
$\beta=54$.}
\label{Sign_Nc8}
\end{figure}               

The most serious limitation of QMC calculations at low temperatures is 
the sign problem. Off half-filling, the sign of
$P(\{s_i\})\propto det [G_c^{\uparrow}(\{s_i\})] \times det [G_c^{\downarrow}(\{s_i\})]$
can be negative so that it can no longer be interpreted as a probability 
distribution. The solution is to reinterpret $\left| P(\{s_i\})\right |$
as the probability of the configuration $\{s_i\}$ and associate its
sign with the measurement.\cite{ELoh_90}  For any operator $O$, this 
becomes
\begin{eqnarray}
\langle O \rangle=\frac{\sum_{s_i}P(\{s_i\}) O({\{s_i\}}) }
{\sum_{s_i}P(\{s_i\})}
=\frac{\sum_{s_i}'\sign(\{s_i\})  O({\{s_i\}})}
{\sum_{s_i}'\sign(\{s_i\})}
\label{sign1}
\end{eqnarray}
where $\sign(\{s_i\})$ is the sign of $P(\{s_i\})$, $O({\{s_i\}})$
is the value of the operator for the field configuration $\{s_i\}$, 
and the primed sums are over configurations generated by importance 
sampling.  In finite system simulations, as the temperature is 
lowered, the average sign becomes exponentially small\cite{ELoh_90,white2} 
so that it is no longer possible to obtain good statistics.  This sign 
problem has posed a formidable challenge in the field of numerical 
simulations for nearly two decades.   

Some recent works have brought some hope. Gubernatis and Zhang\cite{gubernatis} 
and Zhang \cite{zhang} have  shown that by putting a constraint on the fermion 
determinant, one can construct an approximate algorithm which shows some 
improvement on this problem. While the resulting algorithm seems to be free 
from the sign problem, it is possible that the constraint introduced may 
affect the ergodicity of the algorithm.  The ergodicity question is suggested 
by the work of Sorella \cite{sorella} who employ a similar idea as the former 
authors but who arrived at different results.  The most promising new 
direction seems to be that of Chandrasekharan and Weise\cite{chandra}. 
They proposed a new algorithm which is rigorously free from a sign problem
for certain classes of models.  The basic idea is to decompose a configuration 
of fermion world-lines into clusters that contribute independently to the
sign. There are two type of clusters: clusters whose flip changes the sign 
called meron and others that do not modify the sign after a flip. 
Configurations containing meron-clusters contribute 0 to the partition
function, while all other configurations contribute 1.  Hence,  this cluster 
representation describes the partition function as a gas of clusters in the 
zero-meron sector.

\begin{figure}[ht]
\leavevmode\centering\psfig{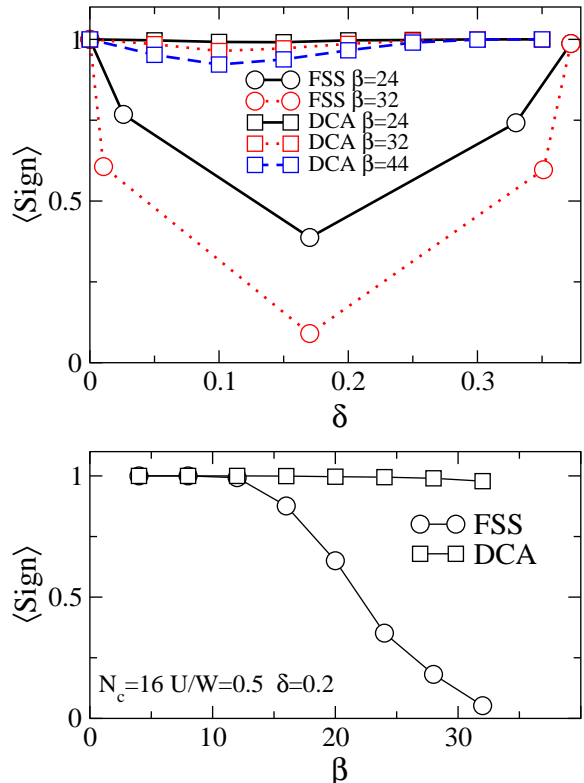}
\caption[a]{A comparison of the average sign for the DCA and 
FS simulations\cite{white2} when $U/W=1/2$, $\delta=0.2$, $t_\perp=0$.}
\label{Sign_Nc16_FSSvsDCA}
\end{figure}

The sign problem remains in DCA simulations, as illustrated in 
Fig.~\ref{Sign_Nc8} where the average sign is plotted versus inverse 
temperature for various values of $U$ when $\delta=0.1$ and $N_c=8$.  
In the inset, the average sign is plotted versus doping when $U=W=2$, 
$\beta=54$ and $N_c=8$.  As in FSS, the sign is worst just off half 
filling.  However, the DCA sign problem is significantly less severe 
than that encountered in FSS.  This is illustrated in 
Fig.~\ref{Sign_Nc16_FSSvsDCA} where the average sign for the DCA and 
the BSS simulations of White {\em{et al.}}\cite{white2} are compared 
when $U/W=1/2$, $\delta=0.2$, and $t_\perp=0$.  When $t_\perp$ is finite, 
the average sign increases further (not illustrated).  We attribute this 
strange behavior to the action of the host on the cluster, but its 
actual mathematical justification is still mysterious.  Nevertheless, 
due to the large reduction in the severity of the sign problem, we are 
able to study the physics at significantly lower temperatures than is 
possible with FSS!

\subsubsection{Single-Particle Properties}

	Much can be learned about the single-particle properties of
the system, especially Fermi liquid formation, from studying the
momentum distribution function $n(\k)$, the single-particle spectra 
$A(\k,\omega)$ and the single-particle self energy $\Sigma(\k,\omega)$.
For a Fermi liquid, the self energy 
$\Sigma(\k_F,\omega) \sim (1-1/Z)\omega - i b \omega^2$ where $b>0$,
$1/Z>1$ and $\k_F$ is a point on the Fermi surface.  The corresponding
$A(\k_F,\omega)$ is expected to display a sharp Lorentzian-like peak,
and $\left|\grad n(\k) \right|$ is also expected to become sharply 
peaked at the Fermi surface.  In each case, these quantities are calculated 
by first interpolating the cluster self energy onto the lattice $\k$ points.

\begin{figure}[htb]
%\epsfxsize=3.3in
%\epsffile{FS.eps}
\leavevmode\centering\psfig{file=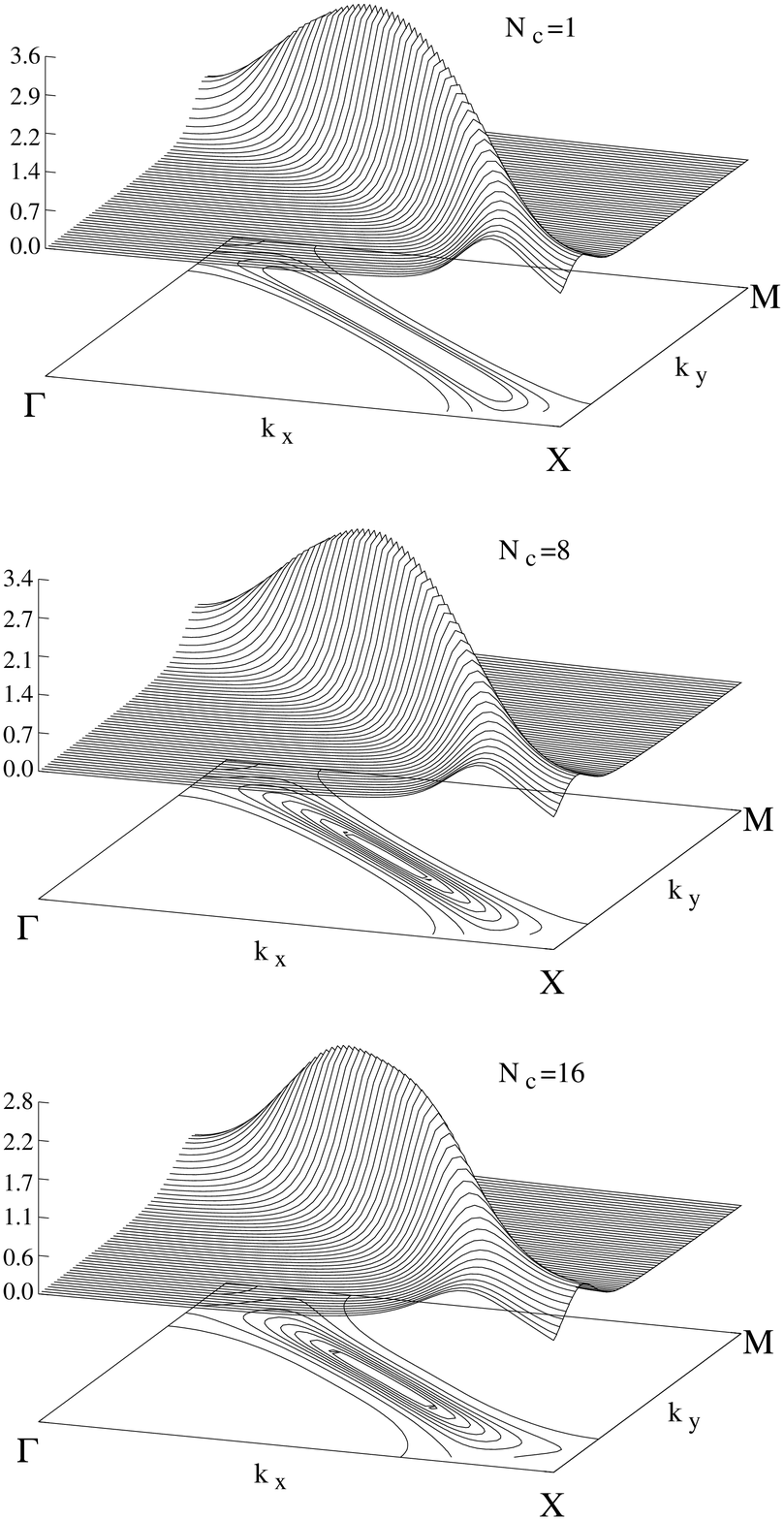,width=3.3in}
\caption{$\left|\grad n(\k)\right|$ versus $\k$ when $U=1$
$\beta=44$, $t_\perp=0$ and $\delta=0.05$ for $N_c=1,8$ and $16$.
}
\label{FS}
\end{figure}

	For example, the gradient of the momentum distribution function
is plotted in Fig.~\ref{FS} when $U=1$, $\beta=44$, $\delta=0.05$ for 
different values of $N_c$ (this temperature would correspond to roughly
room temperature for the cuprates in units where the bare bandwidth 
$W=2$eV).  Apparently, at this temperature, there are
two Fermi surface features, one centered at $\Gamma=(0,0)$ and 
one centered at $M=(\pi,\pi)$.  The Fermi surface centered at
$\Gamma=(0,0)$ has roughly the volume expected of non-interacting 
electrons, so we will call it the electron-like surface and the other
hole-like.  Note that the hole-like Fermi surface becomes more 
prevalent, and the peak near $(\pi/2,\pi/2)$ diminishes, as $N_c$ increases.  
We therefore attribute this behavior to short-ranged correlations.
%********************************************************************
%********************************************************************

\begin{figure}[htb]
%\epsfxsize=3.3in
%\epsffile{Akw.eps}
\leavevmode\centering\psfig{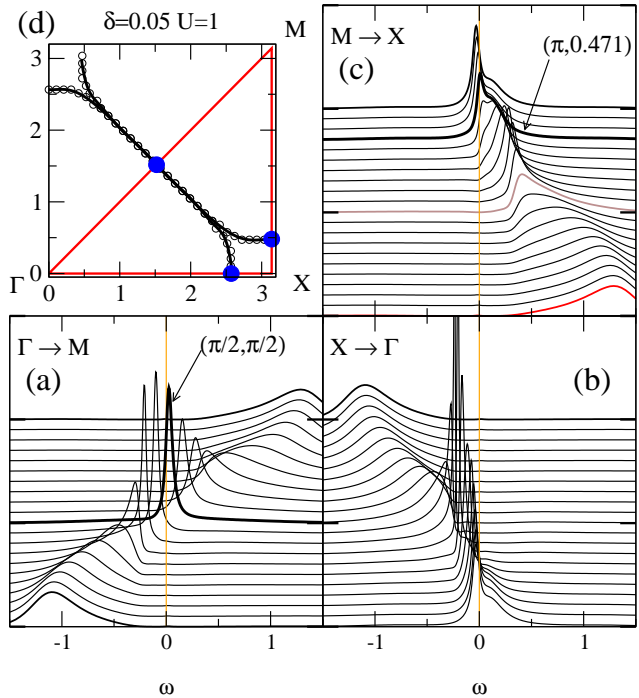}
\caption{ (a)--(c) The single-particle spectrum $A(\k,\omega)$ for $U=1$, 
$\beta=44$, $\delta=0.05$, $t_\perp=0$ and $N_c=16$ along certain high symmetry
directions. The arrows and bold lines in Figs.~(a) and (c) indicate the 
spectra which cross the Fermi energy with a peak closest to $\omega=0$.  
In (b), no such peak is found which crosses the Fermi energy.  (d) the 
maxima of the $\left|\grad n(\k) \right|$ 
data illustrated in Fig.~\ref{FS} plotted versus $\k$.
}
\label{Akw}
\end{figure}

%M= pi,pi
%X= pi,0
	We can further resolve the different surface features, by
investigating the single-particle spectrum $A(\k,\omega)$ as shown
in Fig.~\ref{Akw} for $U=1$, $\beta=44$, $\delta=0.05$ and $N_c=16$.
The graph (d) on the upper left of Fig.~\ref{Akw} plots the corresponding 
location of the maxima of $\left|\grad n(\k) \right|$.  Along the
direction from $\Gamma$ to $M$, $A(\k,\omega)$ shows a relatively
well defined and symmetric peak at $\omega=0$ at the location as 
indicated by the maximum of $\left|\grad n(\k) \right|$.  The only 
notable feature is that the peak is a bit better defined for $\k$ 
closer to the zone center $\Gamma$.  Along the direction from $M$ to 
$X$, the part of the hole-like Fermi surface closest to the $X$ 
point is resolved.   Here the peak in $A(\k,\omega)$ crosses the 
Fermi surface at roughly the same $\k$ where the peak in 
$\left|\grad n(\k) \right|$ is seen; however, the peak in 
$A(\k,\omega)$ is broader and is heavily skewed to higher frequencies.  
Finally, along the direction from $X$ to $\Gamma$, we find very sharp 
peaks in $A(\k,\omega)$; however, none occur at $\omega=0$ indicating
that there are well defined quasiparticle excitations along
this direction, with a small pseudogap, presumably due to the 
short-ranged order.  This pseudogap behavior becomes more pronounced 
as the temperature is lowered.
%********************************************************************
%********************************************************************

\begin{figure}[htb]
%\epsfxsize=3.3in
%\epsffile{doses.eps}
\leavevmode\centering\psfig{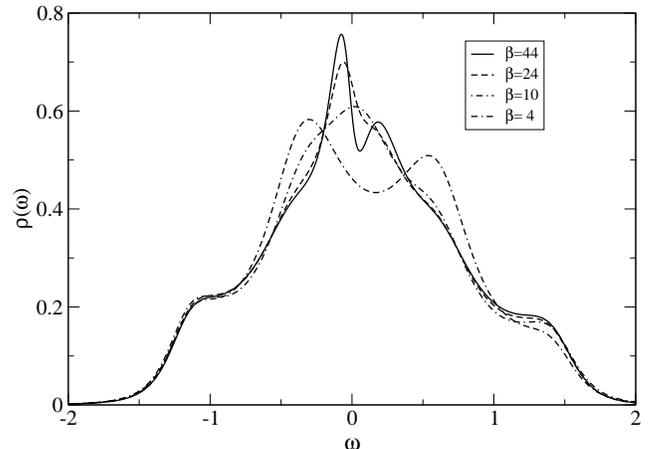}
\caption{The single-particle density of states $\rho(\omega)$ when $U=1$,
$\delta=0.05$, $t_\perp=0$ and $N_c=16$ for several different values of 
the inverse temperature $\beta$.  As the temperature is lowered, 
$\rho(\omega)$ develops a pseudogap due to the short-ranged 
antiferromagnetic order.
}
\label{doses}
\end{figure}

	This can be seen in the density of states, shown in Fig.~\ref{doses}
where the gap is more pronounced.  At high temperatures, $\beta=4$ the 
Hubbard side bands are apparent at $\omega \approx \pm 1/2$.  As the 
temperature is lowered, a central peak begins to develop.  At low 
temperatures, $\beta \agt 24$ a pseudogap begins to develop.
%********************************************************************
%********************************************************************

\begin{figure}[htb]
%\epsfxsize=3.3in
%\epsffile{Sigma_crossings.eps}
\leavevmode\centering\psfig{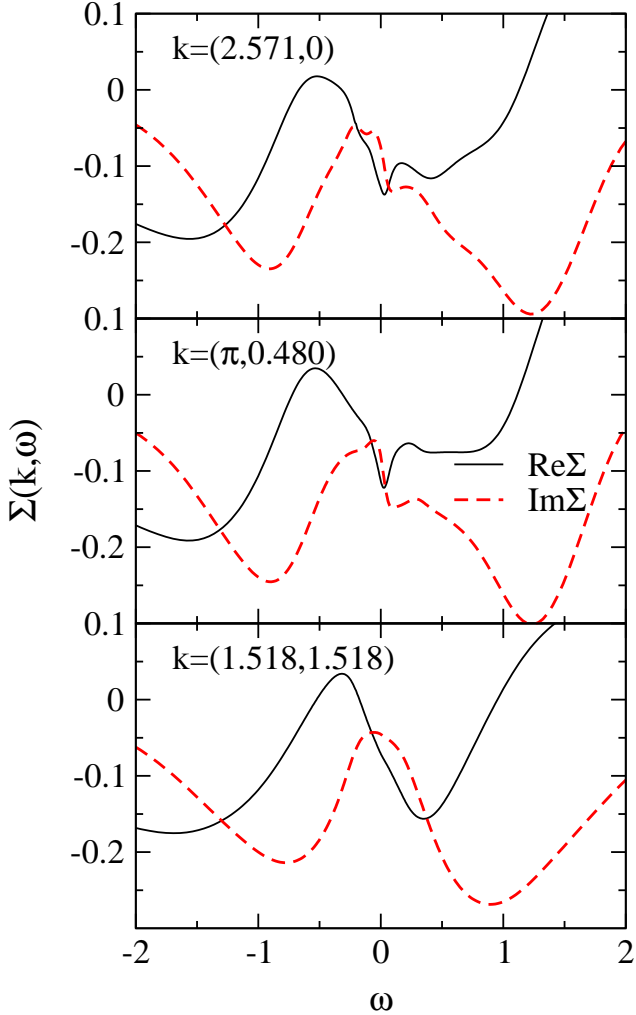}
\caption{The low-frequency self energy, plotted versus $\omega$ for 
the three $\k$-points denoted by filled circles in Fig.~\ref{Akw}(d) 
where the Fermi surface defined by the maxima of $\left|\grad n(\k) \right|$ 
crosses the high symmetry directions.  
}
\label{Sigma_crossings}
\end{figure}

	More can be learned by investigating the self energy directly.  
In Fig.\ref{Sigma_crossings}, both the real and imaginary parts of the 
self energy are plotted for the three values of $\k$ indicated by
filled circles in Fig.~\ref{Akw}(d).  The self energy on the part of 
the Fermi surface along the direction from $\Gamma$ to $M$, looks roughly 
Fermi-liquid like.  However, the self energy on the parts of the Fermi 
surface closest to $X$ have pronounced non-Fermi liquid character,
especially at $\k=(\pi,0.48)$ where the real part displays a minimum
and the imaginary part crosses the Fermi energy almost linearly.
At $\k=(2.571,0)$, the real part again displays a minimum, but the
imaginary part has an almost Fermi-liquid-like maximum at the
Fermi energy, and then once again the scattering rate increases dramatically
at higher energies.  All of the points close to $X$ share this 
dramatic asymmetry; that excitations below the Fermi energy are
much longer lived than those above.  Thus, we expect that the transport
from these parts of the Fermi surface would be predominantly hole-like.
%********************************************************************
%********************************************************************

	The non-Fermi liquid features, including the hole-like distortion
of the Fermi surface, the anisotropy and non-Fermi liquid features of 
the self energy, and the pseudogap in the density of states, become
more pronounced as $N_c$ increases.  Thus, it is reasonable to assume
that these features will persist as $N_c\to \infty$.
                     
\subsubsection{Superconductivity}

\begin{figure}[htb]
%\epsfxsize=3.3in
%\epsffile{Pairfields_zc.eps}
\leavevmode\centering\psfig{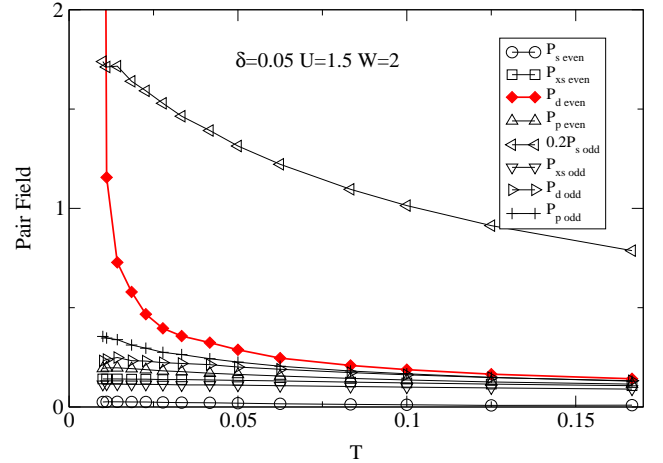}
\caption{Various pair field susceptibilities calculated at the zone
center and plotted versus temperature for $U=1.5$, $\delta=0.05$, 
$t_\perp=0$, and $N_c=8$.  Pairing is found only in the even-frequency 
$\q=0$ d-wave channel. }
\label{Pairfields_zc}
\end{figure}

We searched for many different types of superconductivity, including s,
extended-s, p and d-wave, of both odd and even frequency and we looked
for pairing at both the zone center and corner.  Only the pairing 
channels with zero center of mass momentum (zone center) are enhanced
as the temperature falls.  Of these, only the even-frequency d-wave pair 
field susceptibility diverges. This is illustrated in 
Fig.~\ref{Pairfields_zc} where all of the zone center susceptibilities
are plotted versus temperature for $U=1.5$, $N_c=8$ and $\delta=0.05$.  
%********************************************************************
%********************************************************************

\begin{figure}[htb]
%\epsfxsize=3.3in
%\epsffile{Pd_Nc16.eps}
\leavevmode\centering\psfig{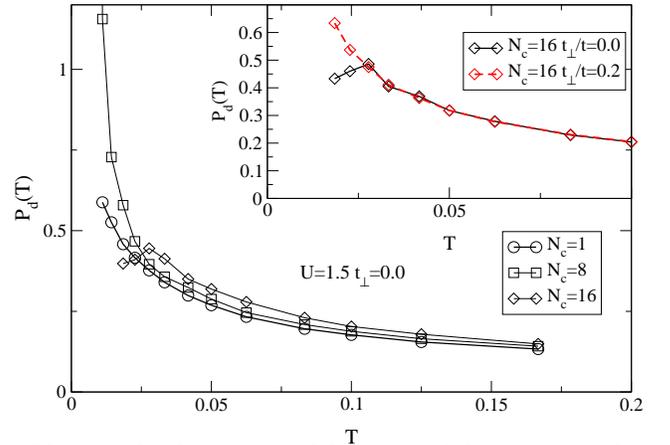}
\caption{The d-wave pair field susceptibility versus $T$ for several 
values of $N_c$ when $U=1.5$, $\delta=0.05$ and $t_\perp/t=0.0$.  In 
the inset the d-wave pair field susceptibility is plotted versus $T$ 
when $N_c=16$, $U=1.5$ and $\delta=0.05$ for $t_\perp/t=0.2$.  }
\label{Pd_Nc16}
\end{figure}

        In contrast to the antiferromagnetic susceptibility which falls 
as $N_c$ increases, the d-wave pair field susceptibility generally rises 
with $N_c$, except at very low temperatures.  This is illustrated in 
Fig.~\ref{Pd_Nc16}.   However, for $N_c=16$ at low $T$, it falls abruptly 
when $T\alt 0.03$.  This behavior is consistent with the lack of 
superconductivity in the purely two-dimensional model.  However, in the 
inset, we see that a very small interplanar coupling $t_\perp/t=0.2$ causes 
the susceptibility to continue to rise with decreasing temperature.  Thus, 
perhaps a very small interplanar coupling is able to stabilize the 
mean-field superconductivity seen in smaller clusters\cite{DCA_Maier2,DCA_Jarrell2}.                                          

%********************************************************************
%********************************************************************
                                                                       
\section{Summary}
\label{Sec_Summary}

We have presented the algorithmic details of the dynamical cluster
approximation. The technique consists of mapping the original
lattice problem to a problem involving a finite cluster dynamically
coupled to an infinite host. The cluster problem may be solved by a 
variety of techniques that include the QMC method, the FLEX
approximation or the NCA. 

An extensive account of a QMC method used to solve the cluster problem 
was given.  Though this algorithm requires significantly greater 
computer power than the Blanckenbecler-Sugar-Scalapino algorithm which 
is often used for finite systems, it has some advantages. First, it 
does not show any numerical instability at low temperature; thus, it 
avoids the time costly matrix factorization step that slows down the 
BSS algorithm.  Second, the algorithm is quite general and can be 
applied to problems that do not have any explicit Hamiltonian 
formulation with known parameters.  Third, a mean-field coupling to 
other degrees of freedom may be easily incorporated.  Fourth, the 
minus sign problem is far less severe in DCA simulations.  This 
allows us to study systems at significantly lower temperatures, with
stronger interactions, or with larger clusters than can be
studied with the BSS algorithm when the sign problem is apparent.

The full DCA algorithm is made of three separate units. In the
first unit the coarse-graining of the lattice is performed and the
resulting self-consistent cluster problem is solved by the QMC 
technique.  This unit requires the formidable computer power 
available on massively parallel computers. The second part
deals with the calculation of the lattice one and two-particle
Green functions from those of the embedded cluster. In the last part 
the analytical continuation of the imaginary-time Green functions 
to real frequency Green functions is performed.

In order to illustrate the originality of the DCA technique, we have 
applied it to the two-dimensional Hubbard model with a small interplanar 
mean-field coupling. 
   
The DCA method was developed to address some of the shortcomings encountered 
in the dynamical mean-field theory. The lack of non-local fluctuations 
in the DMFA leads to incorrect predictions when this method is applied 
to systems in finite dimension. In particular, we have seen that in 
violation of the Mermin-Wagner theorem, the DMFA predicts a 
finite-temperature transition in the two-dimensional Hubbard model. We 
have shown that in the DCA, this transition is progressively suppressed 
as the range of the fluctuations (i.e the cluster size) is increased.  

We also find that a finite-temperature gap persists well into the weak 
coupling regime of the half filled model.  As the DCA systematically 
underestimates the gap formation, these conclusions are valid in the 
limit $N_c\to\infty$.  Since the temperature where the gap opens increases 
with $N_c$, while $T_{\rm{N}}$ decreases, the Slater mechanism is likely not 
responsible for the metal-insulator transition in the two dimensional 
Hubbard model.  The resulting phase diagram is consistent with Anderson's 
view that the effective Hamiltonian for the 2D Hubbard model at 
half-filling for all $U>0$ and $ \Delta \gg T$ (where $\Delta$ is the
gap energy) is the 2D Heisenberg Hamiltonian\cite{anderson}.              

	We find no evidence for a Kondo peak, or the associated
Fermi liquid behavior for the unfrustrated model near half filling.
Since this is an essential feature of the DMFA solution of the
doped model or the half filled model with $U\alt W$, we conclude
that the DMFA is a very poor approximation for the two-dimensional
model, especially for behavior such as the Mott transition, observed
near or at half filling.

When the model is doped, the sign problem becomes significant and will 
certainly affect the quality of the results.  However, the sign problem
is significantly less severe than that found in finite size systems, 
allowing us to explore these model systems at significantly lower 
temperatures, larger coupling or larger clusters than heretofore 
possible.  In the doped model, we find evidence of non-Fermi liquid 
behavior even for relatively small values of $U/W$.  This has been 
observed in the self energy, single particle spectra, and density of 
states.  We also find that the d-wave pair-field susceptibility 
is divergent for small clusters.  A trend that is not present in 
the DMFA because the method cannot treat non-local order parameters.

Finally, the DCA is a very versatile technique that may be applied
to a variety of problems. A straightforward generalization of this
algorithm to the periodic Anderson model in two dimensions will allow 
us to study the physics of the recently discovered two-dimensional 
heavy fermion systems.  In its present form, we have incorporated 
diagonal disorder in the 2D Hubbard model which will allow us to 
address the interesting problem of disorder and interaction in two 
dimensions.   A future improvement of the DCA algorithm itself is to 
insert long range fluctuations in the algorithm which would be treated 
perturbatively.

\noindent{\bf Acknowledgments:} We wish to thank 
H.\  Akhlaghpour,
N.\ Bl\"umer,
A.\ Gonis,
M.\ Hettler,
H.R.\ Krishnamurthy,
E.\ M\"uller-Hartmann,
Th.\ Pruschke,
A.N.\ Tahvildarzadeh,
and
Y.\ Wang
.
We would like to dedicate this paper to E.\ M\"uller-Hartmann, on the 
occasion of his 60th birthday.  
The QMC code described in this manuscript was developed by MJ and CH in 
collaboration with A.N.\ Tahvildarzadeh.    We thank Y. Wang for assistance 
with MPI.  This work was supported by NSF grant DMR-0073308 and by the Ohio 
Supercomputer Center. This research was supported in part by NSF 
cooperative agreement ACI-9619020 through computing resources provided 
by the National Partnership for Advanced Computational Infrastructure 
at the Pittsburgh and San Diego Supercomputer Centers.


\begin{references}
%\addcontentsline{toc}{section}{References}

\bibitem{dagotto}E. Dagotto, Rev.\ Mod.\ Phys.\ {\bf 66}, 763 (1994).  

\bibitem{metzvoll} W.~Metzner and D.~Vollhardt, Phys.\ Rev.\ Lett.\
{\bf 62,} 324 (1989).

\bibitem{muller-hartmann} E.~M\"uller-Hartmann, Z.\ Phys.\ {\bf{B 74}},
507--512 (1989).

\bibitem{pruschke}T.~Pruschke, M.~Jarrell and J.K.~Freericks, Adv.\ 
in Phys.\ {\bf 42,} 187 (1995).

\bibitem{georges}A.~Georges, G.~Kotliar, W.~Krauth and M.J.~Rozenberg,
Rev.\ Mod.\ Phys.\  {\bf 68,} 13 (1996).
 
\bibitem{agonis} The problems encountered in forming cluster corrections
to the DMFA are the same as those encountered in the coherent potential
approximation (CPA) for disordered systems.  For a detailed discussion of 
earlier work on the inclusion of non-local corrections to the CPA see 
A. Gonis, {\em{Green functions for ordered and  disordered systems}}, in 
the series {\em{Studies in Mathematical Physics}} Eds.~E.~van~Groesen 
and E.~M.~DeJager, (North Holland, Amsterdam, 1992).

\bibitem{DCA_Hettler1} M.H.~Hettler, A.N.\ Tahvildar-Zadeh, M.\ Jarrell, 
T.\ Pruschke, and H.R.\ Krishnamurthy, Phys.\ Rev.\ {\bf B 58,} 
7475 (1998).

\bibitem{DCA_Hettler2}  M. H. Hettler,  M. Mukherjee, M. Jarrell, and 
H.R. Krishnamurthy,   Phys. Rev. B 61, 12739 (2000).

\bibitem{DCA_Maier1} Th.\ Maier,  M.\ Jarrell, Th.\ Pruschke and 
J.\ Keller, Eur.\ Phys.\ J.\ B {\bf 13}, 613 (2000).

\bibitem{fye}J.E.~Hirsch and R.M.~Fye, Phys.\ Rev.\ Lett.\ {\bf 56,}
2521 (1986).

\bibitem{FLEX_Bickers}  N.E.\ Bickers, D.J.\ Scalapino, and S.R.\ White,
Phys.\ Rev.\ Lett., {\bf 62}, 961 (1989).

\bibitem{NCA_Bickers}  N.E.\ Bickers, D.L.\ Cox, and J.W.\ Wilkins, Phys.
Rev. B{\bf 36} 2036 (1987).

\bibitem{DCA_Huscroft1} C.\ Huscroft, M.\ Jarrell, Th.\ Maier, 
S.\ Moukouri, A.N.\ Tahvildarzadeh, Phys.\ Rev.\ Lett., {\bf{86}}, 
3691 (2001).

\bibitem{DCA_Moukouri1} S.\ Moukouri, C.\ Huscroft, and M.\ Jarrell, 
to appear in Computer Simulations in Condensed Matter Physics VII, 
edited by D.P.\ Landau, K.K.\  Mon, and H.B.\ Schuttler (Springer-Verlang, 
Heidelberg, Berlin, 2000). 

\bibitem{JARRELLandGUB} M.\ Jarrell and J.E.\ Gubernatis,
Phys.\ Rep.\ {\bf 269,} 135 (1996).

\bibitem{DCA_Moukouri2} S.\ Moukouri and M.\ Jarrell, preprint 
cond-mat/0011247.

\bibitem{treglia} G.\ Treglia, F.\ Ducastelle and D.\ Spanjaard,
Phys.\ Rev.\ {\bf B 21}, 3729 (1980).
 
\bibitem{kuramoto} Y.\ Kuramoto, in {\em{Theory of Heavy Fermions and Valence
Fluctuations}}, edited by T. Kasuya and T. Saso, Springer Ser. Solid State
Sci., Vol. {\bf{62}}, p. 152 (Springer, Berlin, 1985).; C. Kim, Y. Kuramoto 
and T. Kasuya, J. Phys. Soc. Jpn., {\bf{59}} 2414 (1990) .

\bibitem{agd} A.A.\ Abrikosov, L.P.\ Gorkov, I.E.\ Dzyalishinski,
{\em{Methods of Quantum Field Theory in Statistical Physics}},
(Dover, New York, 1975).

\bibitem{DCA_Karan1} K.\ Aryanpour, M.H.\ Hettler and M.\ Jarrell,
in preparation.

\bibitem{peter1} P.G.J.\ van Dongen, Phys.\ Rev.\ B {\bf{50}}, 14016
(1994).

\bibitem{avi} A.\ Schiller and K.\ Ingersent, Phys.\ Rev.\ Lett.\
{\bf{75}}, 113, (1995).

\bibitem{zlatic} V. Zlati\'c and B. Horvati\'c, Sol.\ St.\ Comm.\ {\bf 75},
263 (1990).  

\bibitem{DMFA_Jarrell1} M.\ Jarrell, Phys.\ Rev.\ Lett.\ {\bf 69}, 168(1992)

\bibitem{baym} G.\ Baym and L.P.\ Kadanoff, Phys.\ Rev.\ {\bf{124}}, 
287, (1961).
          
\bibitem{BSS} R.\ Blankenbecler, D.J.\ Scalapino, and R.L.\ Sugar,
Phys.\ Rev.\ {\bf D 24,} 2278 (1981).
 
\bibitem{HHS} J.E.~Hirsch, Phys.\ Rev.\ B {\bf{28}}, 4059 (1983).

\bibitem{NegeleOrland} J.\ W.\ Negele, H.\ Orland, {\em{ Quantum 
Many-Particle Systems}}, (Addison-Wesley Publishing Co.,  1988).

\bibitem{DCA_Maier_comment} T.\ Maier, M.\ Jarrell, and Th.\ Pruschke,
Phys.\ Rev.\ Lett.\  {\bf{86}}, 3691.

\bibitem{white2} S.R.\ White, D.J.\ Scalapino, R.L.\ Sugar, E.Y.\ Loh, 
J.E.\ Gubernatis, and R.T.\ Scalettar, Phys.\ Rev.\ B {\bf{40}}, 506 (1989).

\bibitem{Loh_92} E.\ Y.\ Loh and J.E.\ Gubernatis, in {\it Electronic 
Phase Transitions}, edited by W.\ Hanke and Y.-V.\ Kopaev, (Elsevier, New
York, 1992), chap.\ 4.

\bibitem{jarrell} M.~Jarrell, H.~Akhlaghpour and T.~Pruschke, in
{\it{Quantum Monte Carlo Methods in Condensed Matter Physics}}, 
Eds.\ M.\ Suzuki (World Scientific, Singapore, 1993), p.\ 221--234, 1993.

\bibitem{akima} G.\ Engeln-M\"ullges and F.\ Uhlig, {\em{Numerical 
Recipes with Fortran}}, (Springer, Berlin, 1996).

\bibitem{deisz} J.\ Deisz, private communication.

\bibitem{netlib} See http://www.netlib.org.

\bibitem{numrec_root}{\em{Numerical Recipes in Fortran 77}}, Second 
Edition, W.H.\ Press, S.A.\ Teukolsky, W.T.\ Betterling, and B.P.\ 
Flannery, (Cambridge University Press, 1997), page 364.

\bibitem{tperp} L.B.\ Ioffe and A.J.\ Millis, Science {\bf 285} 1241 (1999);
O.K.\ Andersen {\em{et al.}}, Phys.\ Rev.\ B {\bf 49}, 4145 (1994).

\bibitem{hirsch} J.\ E.\ Hirsch, Phys.\ Rev.\ {\bf 31}, 4403 (1985).

\bibitem{white}  M.\ Vekic and S.R.\ White, Phys.\ Rev.\ {\bf B 47,} 1160 (1993).

\bibitem{lieb} E.H. Lieb and F.Y. Wu, Phys.\ Rev.\ Lett.\ {\bf 20}, 1445 (1968).

\bibitem{caveat_gamma} It is possible that close to the transition
($T\to T_N$) $\gamma$ will return to its mean field value of one.  

\bibitem{ELoh_90} E.Y. Loh, J.E.\ Gubernatis, R.T.\ Scalettar,
S.R.\ White, D.J.\ Scalapino, and R.L.\ Sugar, Phys.\ Rev.\ B
{\bf{41}}, 9301 (1990).

\bibitem{gubernatis} Shiwei Zhang, J.\ Carlson and J.E.\ Gubernatis,
Phys.\ Rev.\ Lett.\ {\bf 74}, 3652 (1995).
 
\bibitem{zhang} Shiwei Zhang, Phys.\ Rev.\ Lett.\ {\bf 83}, 2777 (1999).
 
\bibitem{sorella} S. Sorella, cond-mat/9803107; S. Sorella and
L. Capriotti, cond-mat/9902211.
 
\bibitem{chandra} Shailesh Chandrasekharan and UweJens Wiese,
Phys.\ Rev.\ Lett.\ {\bf 83} 3116 (1999).
 
\bibitem{DCA_Jarrell2} M.~Jarrell, Th.~Maier, M.~H.~Hettler, 
A.N.~Tahvildarzadeh, preprint cond-mat/0011282.

\bibitem{DCA_Maier2} Th.\ Maier, M.\ Jarrell, Th.\ Pruschke, 
J.\ Keller, Phys.\ Rev.\ Lett. 85, 1524 (2000).

\bibitem{anderson} P.W.~Anderson, Phys.\ Rev.\ Lett.\ {\bf 64,} 1839 (1990);
{\bf 65,} 2306 (1990).

\end{references}
\end{document}